\RequirePackage{fix-cm}
\documentclass{svjour3}                     
\smartqed  
\usepackage{rbits}
\usepackage[utf8]{inputenc}
\def\<#1>{\ensuremath{{<}\!\textit{#1}{>}}}
\usepackage{graphicx}
\usepackage{color}
\usepackage{wrapfig}
\usepackage{amssymb}
\usepackage{subcaption}
\usepackage{rotating}
\usepackage{braket}
\usepackage{longtable}
\usepackage{tabularx}
\usepackage{hyperref}
\usepackage{makecell}
\usepackage{adjustbox}
\RequirePackage{rotating}
\newcommand{\acs}[1]{#1}
\newcommand{\acp}[1]{#1}
\newcommand{\dd}{\rm{d}}

\begin{document}

\title{The Lunar Lander Neutron and Dosimetry (LND) Experiment on Chang'E 4
}


\author{Robert F.\,Wimmer-Schweingruber \and
  Jia Yu \and
  Stephan I.\,B\"ottcher \and
  Shenyi Zhang\and
  S\"onke Burmeister\and
  Henning Lohf\and
  Jingnan Guo\and
  Zigong Xu \and
  Bj\"orn Schuster\and
  Lars Seimetz\and
  Johan L.\,Freiherr von Forstner\and
  Ali Ravanbakhsh \and
  Violetta Knierim\and
  Stefan Kolbe\and
  Hauke Woyciechowsky\and
  Shrinivasrao R.\,Kulkarni \and 
  Bin Yuan\and
  Guohong Shen\and
  Chunqing Wang\and
  Zheng Chang\and
  Thomas Berger\and
  Christine E.\,Hellweg\and
  Daniel Matthi\"a\and
  Donghui Hou\and
  Alke Knappmann\and
  Charlotte B\"uschel\and
  Xufeng Hou\and
  Baoguo Ren\and
  Qiang Fu
}

\authorrunning{Wimmer-Schweingruber et al.} 

\institute{R.\,F.\,Wimmer-Schweingruber \at
              Institute of Experimental and Applied Physics, Kiel University, D-24118 Kiel, Germany \\
              Tel.: +49-431-880-3964 
              Fax: +49-431-880-3968 
              \email{wimmer@physik.uni-kiel.de}
              \at
              National Space Science Center, Chinese Academy of Sciences, Beijing, China
           \and
           J.\,Yu \and  S.\,I.\,B\"ottcher \and S.\,Burmeister \and H.\,Lohf  \and  Zigong Xu \and B.\,Schuster \and L.\,Seimetz \and A.\,Ravanbakhsh \and S.\,Kolbe \and V.\,Knierim \and H. Woyciechowsky \and Johan L.\,Freiherr von Forstner\and  Shrinivasrao R. Kulkarni \and A.\,Knappmann \and C.\,B\"uschel 
           \at
           Institute of Experimental and Applied Physics, Kiel University, D-24118 Kiel, Germany 
           \and
           S.\,Zhang \and B.\,Yuan \and G.\,Shen \and C.\,Wang \and Z.\,Chang \at
           National Space Science Center, Chinese Academy of Sciences, Beijing, China
           \and
           J.\,Guo \at
           Institute of Experimental and Applied Physics, Kiel University, D-24118 Kiel, Germany 
           \at
           CAS Key Laboratory of Geospace Environment, Dept.\,of Geophys.\,\&Planet\,Sci., University of Science and Technology of China, Hefei 230026, China
           \at
           CAS Center for Excellence in Comparative Planetology, Hefei 230026, China
           \and
           T.\,Berger \and C.E.\,Hellweg \and D.\,Matthi\"a\at
           Institute of Aerospace Medicine, German Aerospace Center, Linder H\"ohe, D-51147 Cologne, Germany
           \and
           D.\,Hou \at
           University of Chinese Academy of Sciences, Beijing, China
           \and
           X.\,Hou \and B.\,Ren \and Q.\,Fu \at
           China Electronics Technology Group Corporation, NO, 18th Research Institute, Tianjing, China
           \and
           Q.\,Fu \at
           National Astronomical Observatories, Chinese Academy of Sciences, Beijing, China
}

\date{Received: date / Accepted: date}

\maketitle

\begin{abstract}

Chang'E 4 is the first mission to the far side of the Moon and consists of a lander, a rover, and a relay spacecraft. Lander and rover were launched at 18:23 UTC on December 7, 2018 and landed in the von K\'arm\'an crater at 02:26 UTC on January 3, 2019. Here we describe the Lunar Lander Neutron \& Dosimetry experiment (\acs{LND}) which is part of the Chang'E 4 Lander scientific payload. Its chief scientific goal is to obtain first active dosimetric measurements on the surface of the Moon. LND also provides observations of fast neutrons which are a result of the interaction of high-energy particle radiation with the lunar regolith and of their thermalized counterpart, thermal neutrons, which are a sensitive indicator of subsurface water content.

\end{abstract}

\section{Introduction}
\label{sec:intro}


China is currently implementing a series of lunar missions which shall culminate in a crewed station on the lunar surface. The most recent major step in this endeavouring is the challenging Chang'E 4 mission to the far side of the Moon, to the South-Pole Aitken (\acs{SPA}) basin \cite{xu-etal-2018}.  Chang'E 4 landed inside the von K\'arm\'an crater at 177.5991$^\circ$ E, 45.4446$^\circ$ S in selenograpic coordinates at an elevation of -5,935 m \cite{liu-etal-2019}. The \acs{SPA} basin is the largest known impact crater on the Moon and one of the largest in the solar system.  It is clear that a record of such a large impact is key to understanding large impacts in the solar system and thus constrains the impact history in the solar system \cite{potter-etal-2012}.  In addition, it gives insights into the evolution of such large-scale impact basins \cite{head-etal-1993}, and reflects the composition of the lower crust and upper mantle of the Moon.  Inside the \acs{SPA} basin lies the relatively smooth von K\'arm\'an impact crater which is the landing site for Chinese lunar mission Chang'E 4  \cite{ye-etal-2017,wu-etal-2017}.  The coordinates of the landing site and its geological context of the von K\'arm\'an crater and \acs{SPA} basin is discussed in \cite{huang-etal-2018}.

In preparation of human exploration of the Moon and solar system, the Lunar Lander Neutrons and Dosimetry (\acs{LND}) experiment aboard Chang'E 4 is performing the first active dosimetric measurements on the lunar surface.  Radiation is one of the main concerns in human space flight and is a potentially limiting factor for crewed long-term missions such as a mission to the Moon or to Mars \cite{cucinotta-etal-2011}. The complex radiation field of lunar missions is expected to exhibit large temporal variations in the intensity as well as the composition of the radiation and is quite different from that measured on the surface of Earth \cite{gordon2004measurement,haino2004measurements,neher1971cosmic,kremer1999measurements}. Therefore, active dosimetry, which provides time-resolved measurements of the radiation exposure, is a crucial requirement for human spaceflight in general and equally for human exploration of the Moon. 

The \acs{LND} experiment consists of segmented silicon Solid State Detectors (\acp{SSD}) which form a particle telescope that measures the charged particle radiation. A new geometrical arrangement and combination with conversion foils allows \acs{LND} to also measure the electrically neutral component, neutrons and $\gamma$-rays. These contribute a non-negligible fraction to the radiation dose \cite{reitz-etal-2012}. The largest part of the absorbed radiation dose comes from electrically charged galactic cosmic rays (\acs{GCR}), sporadic solar particle events, or - inside the Earth's magnetosphere - from so-called ``trapped'' particles in the radiation belts. On the Moon, the latter is unimportant, but is replaced by an additional, secondary radiation source which is due to the interaction of the \acs{GCR} with the lunar soil (regolith). Apart from charged albedo particles \cite{wilson-etal-2012}, this secondary radiation contains neutral $\gamma$- and neutron radiation. The latter are highly relevant in a radio-biological context \cite{ottolenghi2013assessment}.

Apart from providing information about the temporal variation of the dose rate, \acs{LND} also determines the average quality factor $\langle Q\rangle$ by measuring the Linear Energy Transfer (\acs{LET}) spectrum of the radiation field. For calculating $\langle Q\rangle$ from the measured LET spectrum, the LET dependence of $Q$ as described in ICRP60 and ICRP103 \cite{streffer-2007} is used.\label{page:Q} This $\langle Q\rangle$ can be folded with the measured energy dose $D$ to determine the dose equivalent $H$. Radiation measurement with planar Si-\acp{SSD} telescopes is a well-established method in dosimetry because it is robust and easy to handle (see e.g. \cite{reitz-etal-2005,labrenz-etal-2015,berger-etal-2017}). As just described, it also provides all required information for dosimetric measurements. Its disadvantage is that it can not directly measure the tissue-equivalent dose rate and that its count rate in a non-isotropic radiation field depends on the orientation of the detector. The multiple segments of \acs{LND} allow us to compensate for the latter disadvantage by using the ratios of coincidence count rates from the various segments, very similar to what has recently been done for the radiation field on the Martian surface \cite{wimmer-etal-2015}. 

\begin{figure}
\centering
\includegraphics[width=0.67\textwidth]{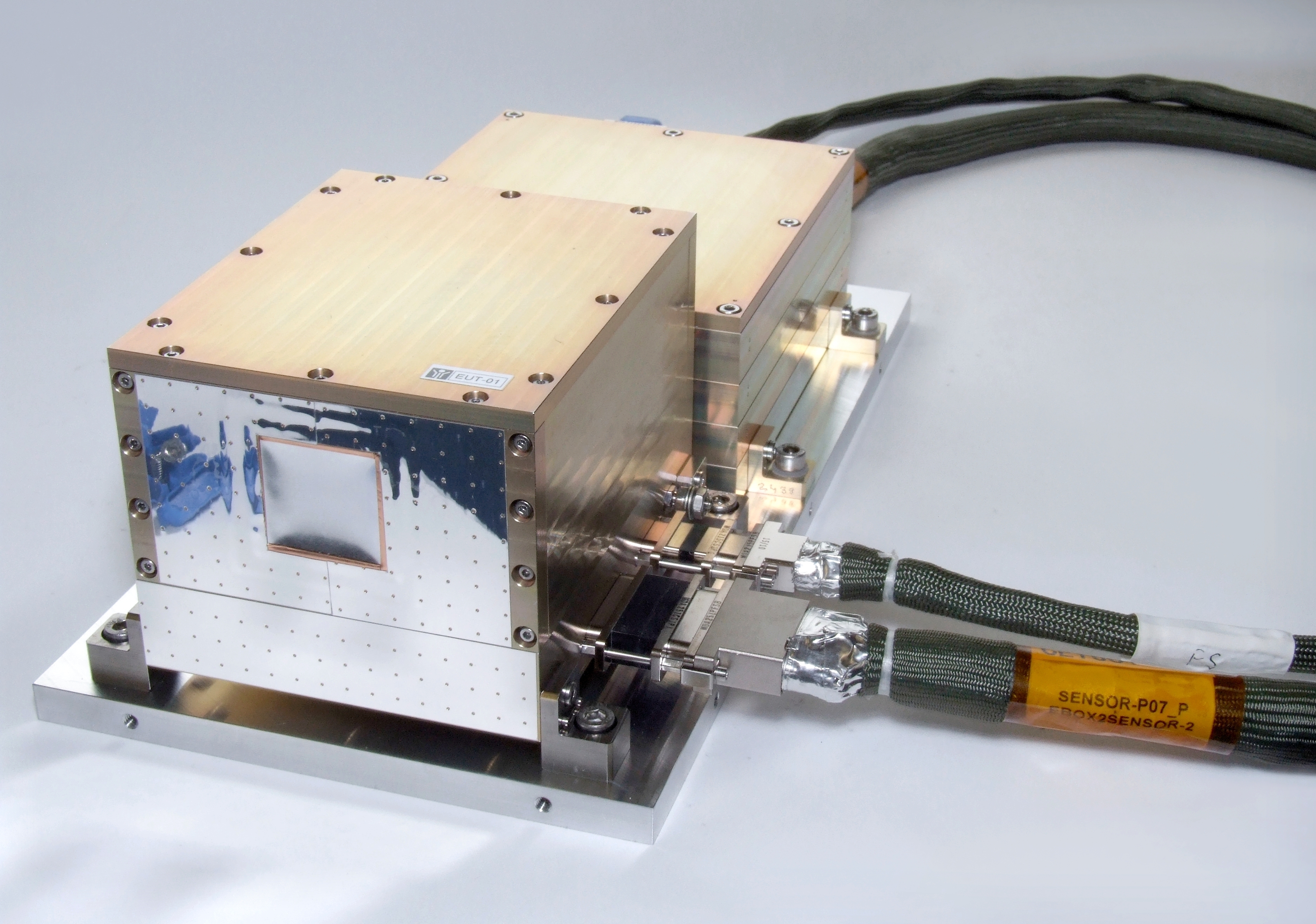}
\caption{{Photograph of \acs{LND} which consists of a sensor head (\acs{SH}, front) and electronics box (\acs{EB}, behind), as well as the power and data harnesses which connect the \acs{SH} with the \acs{EB}. The harness to the the instrument control unit (\acs{ICU}) is not shown.}}
\label{fig:LND}
\end{figure}

Measurements of the radiation field on the lunar surface are crucial to prepare future human exploration of the Moon. \acs{LND} determines the required dosimetric quantities as well as provides measurements of the lunar neutron and $\gamma$-radiation environment.  In this paper we describe the \acs{LND}, starting with its scientific objectives and measurement requirements (Sec.~\ref{subsec:wiss_ziele}), an in-depth discussion of the \acs{LND} design (Sec.\ref{sec:instrument-description}), LND-internal data processing (Sec.~\ref{sec:data-processing}) and data products (Sec.~\ref{sec:data-products}).
Figure~\ref{fig:LND} shows the \acs{LND} sensor head (\acs{SH}) in the front and the \acs{LND} electronics box (\acs{EB}) in the rear.

\section{Science Objectives}
\label{subsec:wiss_ziele}

Despite the aim of landing humans on the Moon in the not too distant future, radiation measurements in the vicinity of the Moon are remarkably scarce. Fairly recent measurements in lunar orbit were provided by the Radiation Dose Monitor (RADOM) on board Chandrayaan-1 \cite{dachev-etal-2011}. The spacecraft reached its operational 100 km circular orbit on November 12, 2008. Measurements showed a dose rate of 0.227 mGy per day averaged over 3545 hours of measurement time (20/11/2008 to 18/5/2009). During the last three months of the mission (20/05/2009 to 28/08/2009) the spacecraft reached a 200 km orbit. The dose rate increased to 0.257 mGy per day owing to the reduction of the lunar shadow effect for cosmic rays and to the increase of the cosmic ray flux related to the reduced solar activity. The minimum of solar activity was reached shortly after the end of the mission in December 2009 and it was not expected that significantly larger radiation exposures due to galactic cosmic rays than the 0.257 mGy per day would occur in these specific orbits \cite{reitz-etal-2012}. This expectation was borne out by measurements in interplanetary space \cite{mewaldt-etal-2010}. Newer measurements have been provided by the Cosmic Ray Telescope for the Effects of Radiation (CRaTER) instrument \cite{spence-etal-2010} on board the Lunar Reconnaissance Orbiter (LRO). CRaTER measured a radiation exposure of about $0.22 - 0.27$ mGy per day in its 50 km orbit \cite{spence-etal-2010b}. Chandrayaan-1 and LRO are the only recent missions which have reported data about the radiation environment of the Moon apart from the Apollo 17 {\em subsurface} measurements of low-energy neutrons by \cite{woolum-etal-1975}. 

In comparison with these meagre orbital data, there is a real dearth of data on the lunar surface. The current knowledge about the radiation environment on the surface of the Moon is based exclusively on calculations using radiation transport models (e.g.\cite{reitz-etal-2012}) with input parameters from models for the galactic cosmic ray spectra and for solar particle events. While such models are based on well-established physical principles, uncertainties in the inputs and interpretations can result in significant differences in the predicted total absorbed dose rate or total dose equivalent rate \cite{matthiae-etal-2017} and come with considerable uncertainties \cite{mrigakshi-etal-2012}. Measurements of the lunar neutron density at depths of 20 - 400 g/cm$^2$ within the lunar subsurface were performed during the Apollo 17 mission \cite{woolum-etal-1975}. These measurements were performed to determine the depth profile of neutrons in the lunar regolith. This is an important factor if one aims to determine time scales of the lunar surface mixing processes \cite{eugster-etal-1970}. These authors stressed the importance of measuring the spectrum of neutrons at the lunar surface, but to our knowledge, this has never been done.

The high biological effectiveness of neutrons has been studied by multiple authors (e.g. \cite{ottolenghi2013assessment,streffer-2007,Baiocco2016}). This is due to the fact that neutrons can easily interact with nuclei in biological tissue and produce secondary charged particles. These low-energy short-range particles interact with the tissue and exhibit a high ionization density and therefore high potential to damage living cells by the well-known mechanisms such as direct DNA damage or production of free radicals. Thus, determination of the neutron spectrum at the lunar surface is a crucial measurement in preparation of future human exploration of the Moon.

Thus, LND has the following two major science objectives:

\begin{enumerate}
\item {\bf Dosimetry for human exploration of the Moon:} Determination of time series of dose rate and of \acs{LET} spectra in the complex radiation field of the lunar surface: Radiation is one of the largest long-term risks in manned spaceflight. \acs{LND} will measure the dose for a manned lunar mission as well as the temporal variation of the dose rate. For the interpretation of these dosimetric data, the quality factor, $Q$, needs to be known. This is derived from the LET spectra measured by LND as described in paragraph 2 on page~\pageref{page:Q}. A novel contribution of \acs{LND} is to also provide measurements of the neutral (neutron and $\gamma$-ray) radiation which is an important contribution to the radiation dose. Measuring the neutron spectrum at the surface also provides important information to understand lunar surface mixing processes. 

\item {\bf Contribution to heliospheric science:} Determination of particle fluxes and their temporal variations on the far side of the Moon: There are already a number of spacecraft close to Earth (especially at Lagrange point L1) that measure the heliospheric particle radiation environment in the vicinity of Earth. Such a heliospheric "cluster" of spacecraft is augmented by \acs{LND} because it provides an additional, and displaced measurement location. This new data can contribute to the understanding of particle propagation and transport in the heliosphere through measurements of the onset times of particle events and the timing of changes in the energetic particle flux. This is an active field of research in heliophysics and \acs{LND} can contribute to it by providing such data. \acs{LND} provides data at high cadence (up to one minute) to allow precise determination of the onset times. Solar particle events have different spectra than the \acs{GCR} that do not extend to as high energies, but can nevertheless be a danger for astronauts on the surface of or orbiting the Moon if their high-energy flux is sufficiently high (e.g. \cite{guo2018generalized}). 

\end{enumerate}

In addition to the two aforementioned major science objectives, \acs{LND} also has a technological demonstration objective. The extremely large cross section of gadolinium (Gd) to capture thermal neutrons and subsequently emit a conversion electron has been exploited for thermal neutron measurements on Earth, but to our knowledge never before in space (See Sec.~\ref{sec:instrument-description} for an in-depth discussion.). Thermal and epithermal neutrons serve as sensitive probes for the shallow subsurface water (proton) \cite{feldman-etal-1998-a,feldman-etal-1998-b} and FeO content \cite{elphic-etal-1998}. \acs{LND}'s technological objective is thus to demonstrate the technological readiness level of this detection technique for space-based exploration missions. Because of the long and cold lunar nights, Chang'E 4 has to rely on radioisotope thermoelectric generators (\acp{RTG}) and several radioactive heater units (\acp{RHU}) on the lander and rover, and so the scientific success of this method can not be ascertained at this point. The following two science questions illustrate the potential of this measurement technique. 

\begin{enumerate}
\item {\bf Determine the subsurface water content in the South-Pole Aitken Basin:} Subsurface water on the Moon is a precious resource which would be needed for long-term human presence of humans on the Moon. It is thus critical to understand the amount of water on the Moon. Because the South-Pole Aitken Basin has some permanently shadowed regions, it is believed that it may harbor significant amounts of water \cite{feldman-etal-2001}. \acs{LND} is sensitive to the subsurface water content and determine the flux of thermal neutrons averaged over a small area beneath the lander and a larger area surrounding the lander. See section~\ref{sec:instrument-description} for a more thorough discussion. At the time of writing this paper, the influence of the \acs{RTG} and \acp{RHU} on this measurement is still being assessed.

\item {\bf Determine the FeO content in the South-Pole Aitken Basin:} The South-Pole Aitken Basin is the second largest known impact crater in the solar system. It is therefore also a ``deep hole'' in the crust of the Moon and allows us to peer into mantle material of the Moon. Measurements of Lunar Prospector and Clementine show different results for the abundance of heavy elements (which are dominated by FeO) \cite{lawrence-etal-2002} which is puzzling, but also of high importance to understanding the origin of the Moon and thus the solar system.
\end{enumerate}

To achieve the science objectives listed above, \acs{LND} is designed to perform several measurements of the lunar radiation environment, which are discussed in more detail in Sec.~\ref{sec:data-products} and summarized here:
\label{sec:measurement-requirements}

%
\begin{enumerate}
\item Time series of the charged and neutral particle dose rate in Si at a cadence of up to 1 minute.
\item Coarse \acs{LET}-spectra at a cadence of up to 1 minute.
\item Coarse charged particle spectra at a cadence of up to 1 minute.
\item Coarse energy deposit spectrum of neutral particles at a cadence of up to 1 minute.
\item Count rates of thermal neutrons at a cadence of 10 minutes.
\item High-resolution measurements of all these quantities as well as heavy ions at a cadence of 1 hour.
\end{enumerate}
%

\section{Instrument Description}
\label{sec:instrument-description}

\acs{LND} is mounted inside the $-Y$ payload compartment of the Chang'E 4 lander and consists of a Sensor Head (\acs{SH}) and an Electronics Box (\acs{EB}) (see Fig.~\ref{fig:LND}). The LND \acs{EB} connects to the Instrument Control Unit (\acs{ICU}, not shown) which serves as the electrical and data interface of \acs{LND} with the lander. 
The sensor head is mounted on a bracket close to the upper deck of the lander which has a re-closeable opening which allows an unobstructed field of view (\acs{FOV}) into the sky. This opening is closed during lunar nights to conserve heat inside the payload compartment. The \acs{LND} \acs{FOV} points approximately 13$^\circ$ from the direction of the normal to the lander deck or along ($\vec{n} = (0.9591, 0.0644, 0.2772)$ in (sky, west, south) coordinates.). The small detector signals are pre-amplified, shaped, and converted to digital signals inside the \acs{SH}. They are driven to the \acs{EB} via \acs{UART}/\acs{LVDS}\footnote{Universal Asynchronous Receiver/Transmitter, Low voltage Differential Signaling} where further analysis of the signals is performed. \acs{SH} and \acs{EB} are connected by two harnesses, one for power supply, heater, thermistor and bias voltage of detectors, and the other one for \acs{ADC}s' control signals and data readout (see Fig.~\ref{fig:LND}).  The design with a separate sensor head and electronics box was driven by the requirement to keep the sensor head as cool as possible in the challenging thermal environment of the Chang'E4 lander.

\begin{figure}
\centering
\includegraphics[width=1\textwidth]{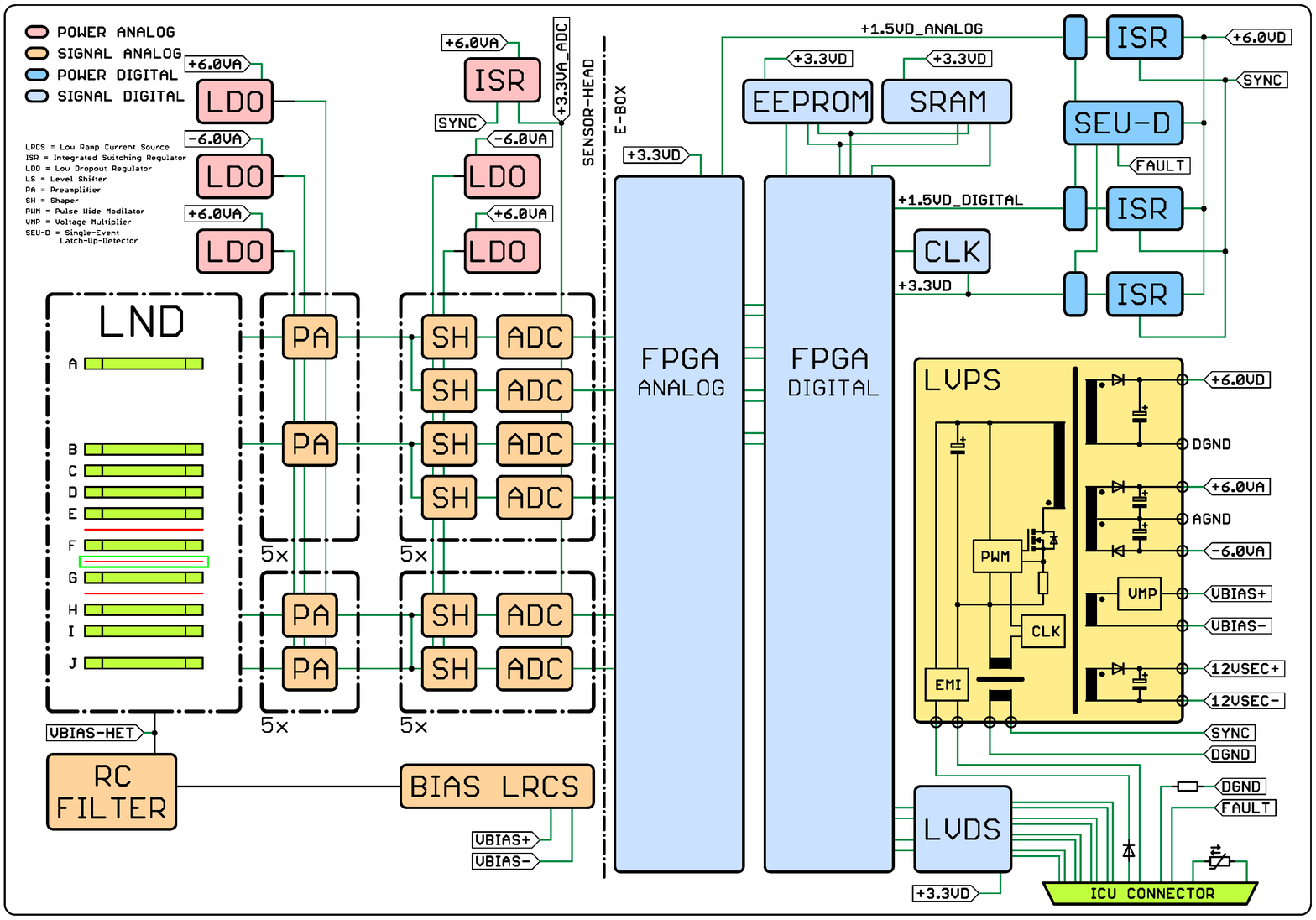}
\caption{Functional block diagram of \acs{LND}. All 10 detectors (indicated by A - J inside the dash-dotted rectangle marked LND) and their front-end electronics (pre-amplifiers (PA), shapers (SH), and analog-to-digital converters (ADCs)) are accomodated in the sensor head (left half of the schematic). All other electronics such as the analog and digital field programmable gate arrays (FPGAs) low-voltage power supply (LVPS) and  are housed in the electronics box (E-Box, to the right of the dash-dotted vertical line).}
\label{fig:block_diagram}
\end{figure}

Figure~\ref{fig:block_diagram} shows a functional block diagram of \acs{LND}. The left-hand side of it (divided by the vertical dashed line) shows the functions performed in the \acs{LND} \acs{SH}, the right-hand side the \acs{LND} \acs{EB} functions.  The \acs{LND} detector system, pre-amplifiers, shapers and analog to digital converters (\acs{ADC}s) are all contained in the \acs{LND} \acs{SH}. Twenty charge sensitive pre-amplifiers are integrated onto a printed-circuit board (\acs{PCB}) which amplifies trigger signals generated by the \acs{LND} detector system.  The pre-amplifiers are followed by 30 pairs of shaper and \acs{ADC} (analog to digital conversion) circuits which shape and and then sample the amplified signals before 
they are sent to the \acs{LND} \acs{EB} via the harness. The digitized values from the \acs{SH} are pulse-height analyzed (\acs{PHA}) by the analog \acs{FPGA} (Field-Programmable Gate Array) in the \acs{EB}. The analog \acs{FPGA} also controls the \acs{ADC}s, performs digital filtering, the front-end trigger processing, and housekeeping data acquisition. The data from the analog \acs{FPGA} are sent to the digital \acs{FPGA} for additional processing and storage before being sent to the \acs{ICU}.  The additional processing in the \acs{FPGA}s is discussed in Sec.~\ref{sec:data-processing}.
The \acs{LND} \acs{EB} also contains the \acs{LVPS} board, which supplies latch-up protected power for \acs{LND} as well as the bias voltage for the \acs{LND} \acs{SSD}s. 

\subsection{Mechanical and Thermal Design Considerations}

\begin{figure}
  \centering
  \includegraphics[width=0.7\textwidth]{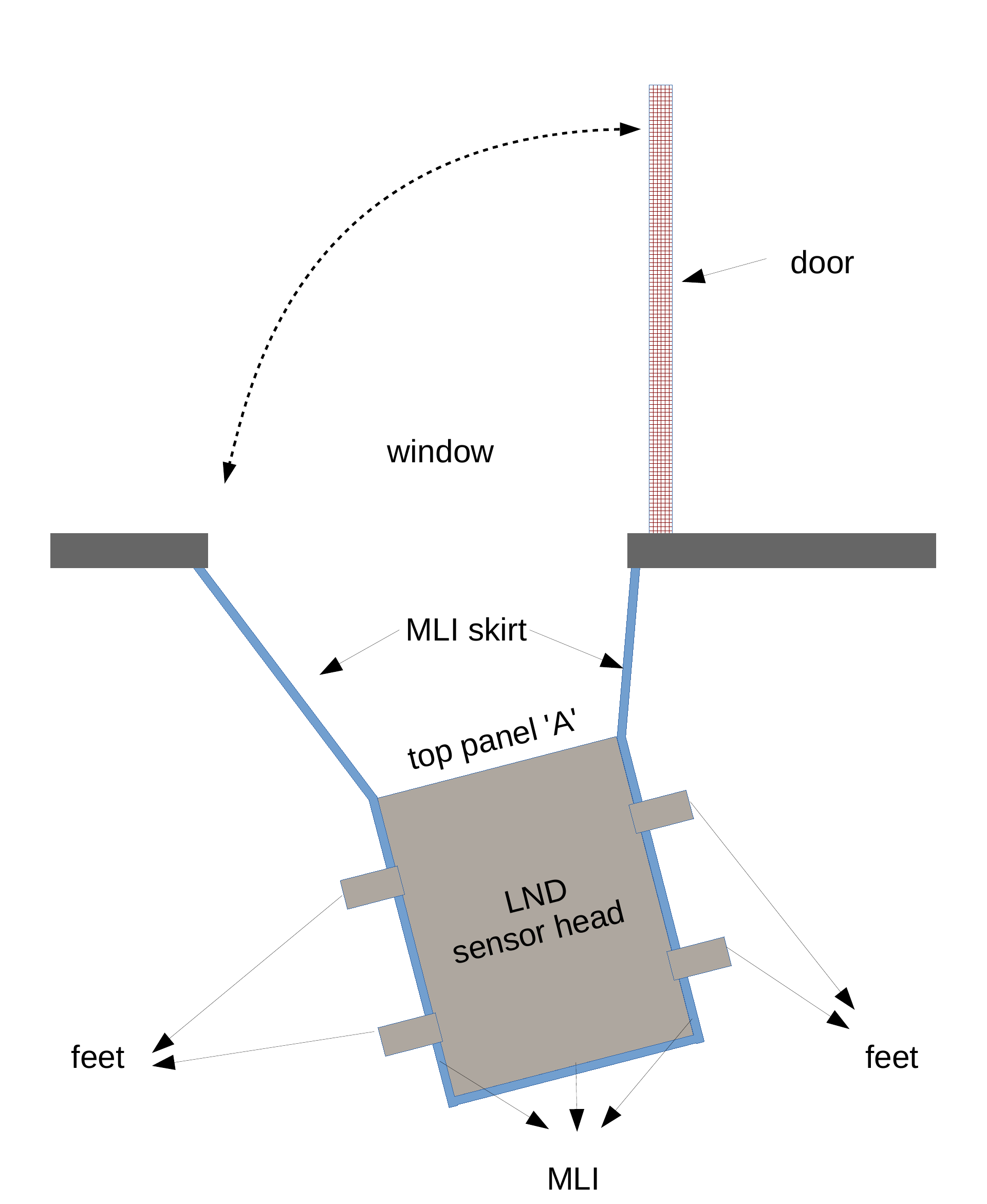}
  \caption{Schematic view of the \acs{LND} sensor head in the payload compartment of the Chang'E 4 lander. \acs{LND} is tilted by 13 degrees with respect to the top panel normal, as indicated in this sketch. There is no tilt in the other direction. }
  \label{fig:sensor head and top window}
\end{figure}


As already mentioned, \acs{LND} is mounted in the $-Y$ compartment of Chang'E 4 Lander and is tilted by 13 degrees, as shown in Fig.~\ref{fig:sensor head and top window}. The sensor head sits on a bracket in order to achieve a unobstructed FOV through the top window of $-Y$ compartment. Additionally, the position of the sensor head is very close to the window not only for the purpose of unobstructed FOV but also to radiate as much heat to the sky as possible via the second surface mirror tape with a reflectivity to absorptivity ratio of $\sim 8$ which is applied to the front panel of the \acs{LND} \acs{SH}. This keeps the sensor head cool and reduces the importance of leakage current as a noise source.  The size of the top window is 16cm $\times$ 23cm and is open during lunar day time and closed during the cold lunar night.  The \acs{LND} \acs{SH} is further insulated from the warm payload compartment by a ``skirt'' of MLI (10 layers, emissivity 0.035) which is also shown in Fig.~\ref{fig:sensor head and top window}. This connects from the top of the window and completely surrounds the \acs{LND} \acs{SH}, effectively thermally decoupling \acs{LND} \acs{SH} from the payload compartment. 

The temperature at the interface of the \acs{LND} bracket and its mounting panel of the lander was modeled to vary from -20$^{\circ}$C to +55$^{\circ}$C in the operational case and -60$^{\circ}$C to +70$^{\circ}$C in the non-operational case. The interface temperature of the \acs{LND} electronics box and the lander is expected to lie between -20$^{\circ}$C and +55$^{\circ}$C in the operational case and -70$^{\circ}$C to +70$^{\circ}$C in the non-operational case. During the first lunar day, temperatures varied between -7.97$^{\circ}$C and 48.10$^{\circ}$C, with the highest temperatures occurring on the sensor head analog board, and the lowest in electronics box.

\subsection{The \acs{LND} Sensor Head}
\label{sec:LND-SH}

This section gives an overview of the \acs{LND} sensor head, more detailed descriptions of the different arrangements are given farther down in this paper, as are several of the figures which are referred to here for the first time. 

\begin{figure}
  \centering
  \includegraphics[width=0.6\textwidth]{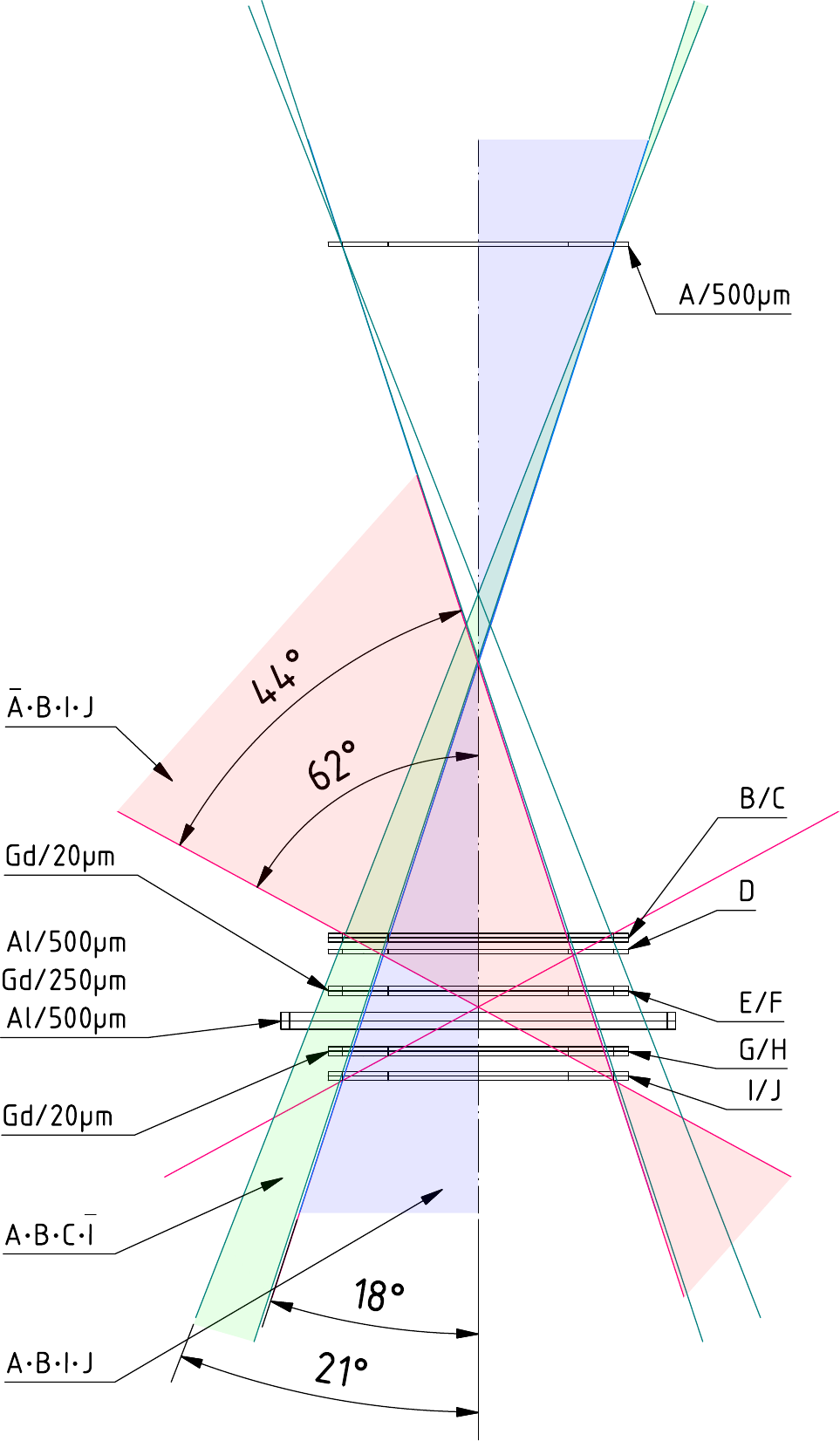}
  \caption{The \acs{LND} telescope consists of 10 segmented 500 $\mu$m-thick Si solid-state detectors (SSDs), A - J. Detectors B/C, E/F, G/H, and I/J are mounted in a sandwich configuration, the inner segments of the A and B detectors, A1 \& B1, span the telescope opening. See text for a more detailed discussion. Angles are measured from the middle to middle of the detector edges, as discussed on p.~\pageref{page:angles}.}
  \label{fig:LND-detector-stack}
\end{figure}
\begin{figure}
  \centering
  \includegraphics[width=0.67\textwidth]{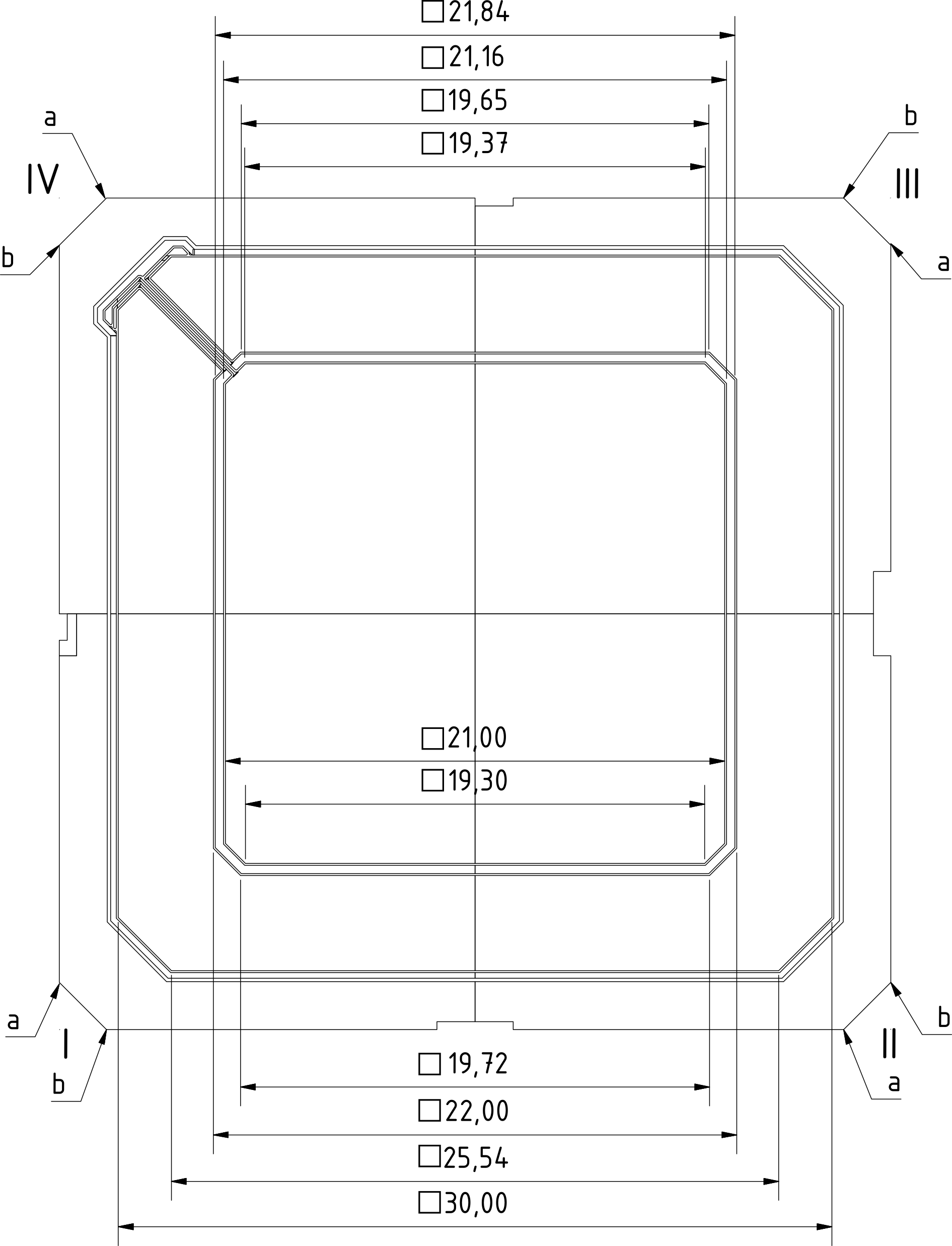}
\caption{LND detector layout shows the sensitive areas of LND's detectors. The inner segments are referred to as ``1'', the outer segments as ``2''. Dimensions are in mm. The area of the inner segment is 440 mm$^2$. The outer segment and cross-talk ring are connected and their area adds up to 435.6 mm$^2$, uncertainties in the measurements in this Figure are less than 0.01 mm.}
\label{fig:detector}
\end{figure}

The \acs{LND} \acs{SH} consists of a stack of ten Si solid-state detectors (\acs{SSD}s) each with a nominal thickness\footnote{Their measured thicknesses are given in Tab.~\ref{tab:LND-segments}} of 500 $\mu$m. They are arranged in a charged-particle telescope configuration as shown in Fig.~\ref{fig:LND-detector-stack} in which the \acs{SSD}s are labeled A through J. The measured thicknesses of the individual detectors are given in Tab.~\ref{tab:LND-segments}. Each detector is segmented into an inner and outer segment with approximately the same areas, as shown in Fig.~\ref{fig:detector}. This telescope configuration is used to detect charged particle radiation and identify its composition. Fast neutrons and $\gamma$-radiation are detected in the inner segment of the C detector in anti-coincidence with all other detector segments. A 20 $\mu$m thin Gd foil is sandwiched between detectors E\&F and G\&H and is used to detect thermal neutrons. A 250 $\mu$m thick Gd foils is sandwiched between two 500 $\mu$m Al plates (shown in green) to shield the upward (downward) thermal neutron flux from the E\&F (G\&H) detectors and thus allow directional information, as discussed below in Sec.~\ref{sec:neutral}.

Fig.~\ref{fig:detector} shows the detector segments and detailed dimensions. 

\subsection{Detectors, Segmentation, and Gain Stages}

As already discussed, the detectors on \acs{LND} are all segmented, some are read out with two energy gains. Table~\ref{tab:LND-segments} gives the conversion factors from \acs{ADC} value to energy (in keV) for \acs{LND}'s detector segments. For instance, detector segment A1 is read out at Low gain (A1L) and High gain (A1H), while the two segments of detector E, E1 and E2, are both read out with one gain. This information is only needed when analyzing \acs{PHA} data. The conversion has already been applied to the published \acs{LND} data. 

\begin{table}
\centering
\begin{tabular}{|lrrr|}\hline\hline
detector/	&thickness	&gain 1	& gain 2\\
segment &    [$\mu$m]        &[keV/mV]  & [keV/mV] \\\hline
A1	   & 514	        & 12.425          & 186.476 \\
A2	   & 	                & 12.530          & 183.837 \\
B1	   & 509	        & 12.447          & 179.555 \\
B2	   &	                & 12.800          & 199.022 \\
C1	   & 505	        & 12.613          & 159.538 \\
C2	   & 	                & 12.400          & 178.805 \\
D1	   & 508	        & 12.322          & 182.632 \\
D2 	   &	                & 12.480          & 183.416 \\
E1	   & 502	        & 4.037	        & --\\
E2	   &	                & 4.142	        & --\\
F1	   & 504	        & 3.993	        & --\\
F2	   &	                & 4.051	        & --\\
G1	   & 514	        & 4.073	        & --\\
G2	   &	                & 4.070	        & --\\
H1	   & 502	        & 4.062	        & --\\
H2	   &	                & 4.065	        & --\\
I1	   & 505	        & 12.403        & 163.653 \\
I2	   &	                & 12.144        & 158.402 \\
J1	   & 509	        & 3.990	        & --\\
J2 	   &	                & 4.025	        & --\\\hline\hline

\end{tabular}
\caption{LND detector segments, measured detector thicknesses and gains. }
\label{tab:LND-segments}
\end{table}


The geometry factors of \acs{LND} are defined by coincident measurements in A1 and B1 up to a depth of detector C. For signals in detectors D -- J, a coincident measurement in A1 and D1 is required, as well as a larger energy deposition in D1 than in D2. This results in a geometry factor of 0.58 cm$^2$sr for A1\&B1 and 0.55 cm$^2$sr for A1\&D1 with uncertainties less than 1\%.\label{page:geofac} The geometric factors of \acs{LND} were calculated using GEANT4 \cite{agostinelli-etal-2003} for an isotropic flux of particles, i.e., from $4\pi$. $4 \pi$ geometry factors are given here because there is also a (small) "return" flux of secondary particles from the Moon. 

For dosimetric quantities shielding by structures in the lander, etc.\,is not that important. This allows a larger geometric factor (22.23 cm$^2$sr) by using the combination of the B (B1\&B2) and I (I1\&I2) detectors which provides high counting statistics. Again, the geometry factor given here is calculated for a $4 \pi$ radiation field.


\subsubsection{Charged-Particle Telescope}
\label{sec:telescope}

A schematic view of the \acs{LND} \acs{SH} charged-particle telescope is shown in Fig.~\ref{fig:LND-detector-stack}. It can be divided into an upper and lower half. The upper four detectors (A--D) form a detector arrangement similar to the Flight Radiation Environment Detector (\acs{FRED}) \cite{moeller-etal-2011,moeller-etal-2013,moeller-etal-2013b} which was based on the Ionizing RAdiation Sensor (\acs{IRAS}\footnote{ESA's original Exomars mission included the Humboldt lander with its Pasteur payload. The Pasteur payload included the Ionizing Radiation Assessment Sensor (\acs{IRAS}, \cite{wimmer-etal-2008}), and parts of is design are used in \acs{LND}.  After assembly of the \acs{IRAS} prototype, it was reused as \acs{FRED} in several high-altitude balloon flights \cite{moeller-etal-2013,moeller-etal-2013b}  and was also qualified for use in civil aircraft \cite{moeller-2013}. \acs{FRED} used an IRAS-telescope of four 300 $\mu$m Si-\acs{SSD}s and was calibrated with neutrons at PTB and iThemba \cite{moeller-2013,moeller-etal-2013b}}.) which was developed as part of \acs{ESA}'s Exomars program. The lower half of \acs{LND} (detectors E--J) consists of a closely packed stack of six more Passivated Implanted Planar Silicon (\acs{PIPS}) detectors and an Al-Gd-Al absorber for thermal neutrons. Two pairs (E\&F and G\&H) clamp a 20 $\mu$m thin Gd foil, while I\&J complete the charged-particle telescope. 

The full opening angle of 29.4$^\circ$ of the \acs{LND} FoV is determined by the inner segments of detectors A and B (A1 and B1) and the outer segment of detector J (J2) as sketched in Fig.~\ref{fig:LND-detector-stack}. Note that this angel is not shown in Fig.~\ref{fig:LND-detector-stack}. The opening angles (measured from the middle to the middle of the detector edges) and geometric factors of relevant detector combinations are given above. \label{page:angles}

This arrangement allows to determine which ions stop in the various detectors and gives a good estimate of the penetrating particle flux. The resulting rather longish telescope ensures that the full energy deposit of a particle which triggers B1 and A1 is measured for stopping ions or that it is identified as a penetrating ion. The primary energy ranges for ions stopping in B--I are given in Tab.~\ref{tab:stop-energies}. For dosimetric measurements discussed in subsection~\ref{sec:dosi} a higher count rate is advantageous and both inner and outer segments of the relevant detectors are used. Because electrons scatter much more than ions, this telescope configuration is less sensitive to their primary energy, although some information is still retained as can be seen in Fig.~\ref{fig:electrons} and discussed there.

The telescope aperture is covered by a layer of thin aluminized Kapton foil (50.8 $\mu$m Kapton and 25.4 $\mu$m of Al). The size of this window corresponds to the opening of the telescope spanned by the A1$\cdot$B1 detector segments projected onto the front housing surface of the \acs{LND} \acs{SH}.  The energy which particles loose in this protective foil is small compared to the energy loss in the front detector, A, and has been accounted for in the values of energy ranges given in Tab.~\ref{tab:stop-energies}. 

\acs{LND} can measure the composition of the charged particles using the combination of the energy deposited in the front detector (A) and the total deposited energy in the (B - I) detector stack. This allows \acs{LND} to measure the composition of the stopping particles, which is important for dosimetric purposes but also for heliospheric science. \acs{LND} provides energy spectra up to 30 MeV/nuc, i.e., above the often-seen ``knee'' in the interplanetary energy spectrum. This ``knee'' is an important indicator of the acceleration process \cite{reames-2014}.  Of course the majority of all particles will penetrate \acs{LND}, nevertheless, their energy deposit remains proportional to the square of their nuclear charge and, therefore, \acs{LND} measures and discriminates all major minimally ionizing particle species. Knowledge of the particle composition is important for dosimetry and helps to better understand the \acs{LET} spectrum (See Sec.~\ref{sec:dosi}).

\subsection{Neutral Particle Detection}
\label{sec:neutral}

\begin{figure}
\centering
\begin{subfigure}{.495\textwidth}
  \centering
  \includegraphics[width=.9\textwidth,clip=]{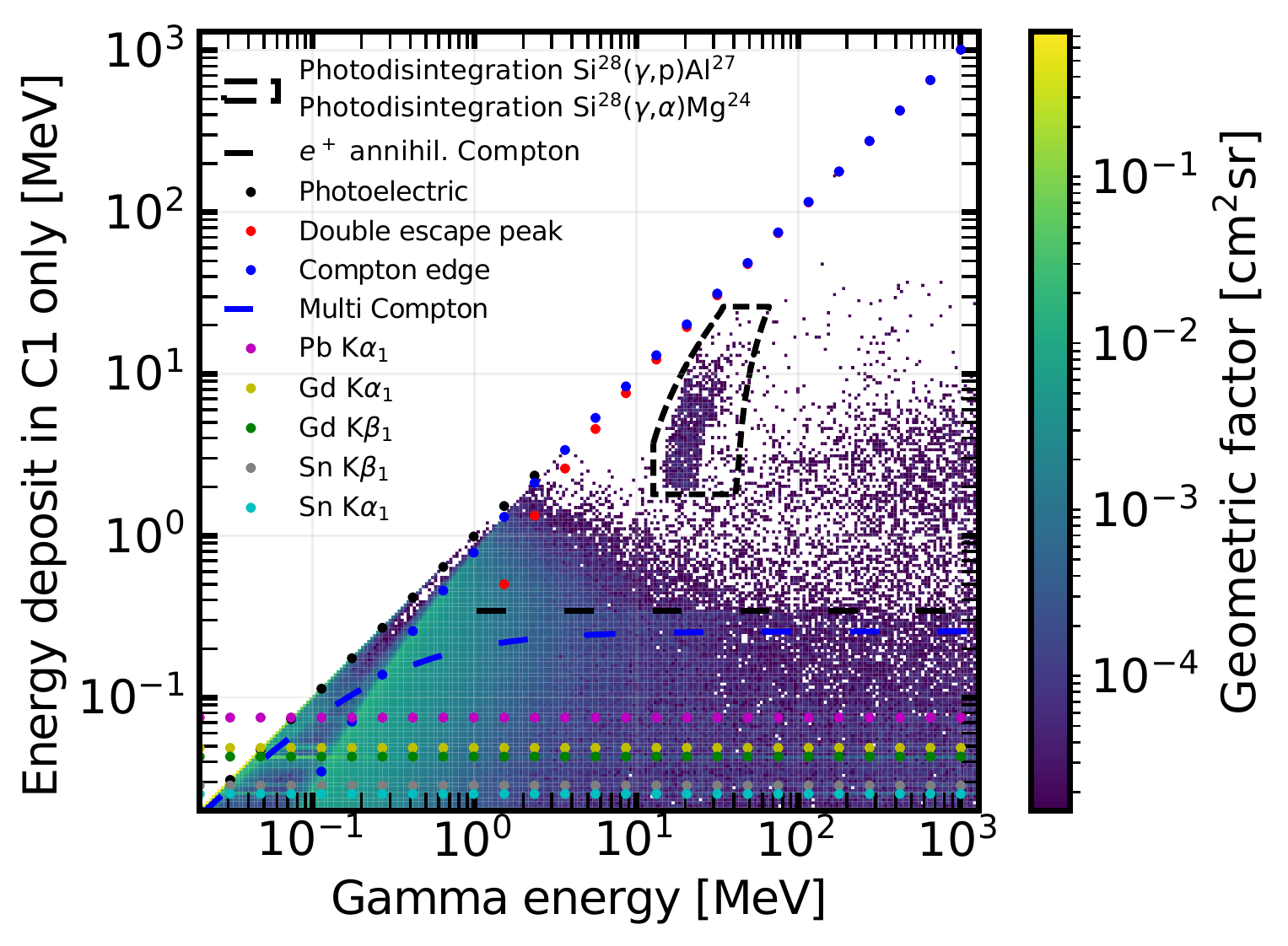}
  \caption{Geometry factor for the detection of gamma rays. Colored dotted lines indicate various interaction processes discussed in the text.}
  \label{fig:gamma_geofac}
\end{subfigure}\hfill%
\begin{subfigure}{.495\textwidth}
  \centering
  \includegraphics[width=.9\textwidth,clip=]{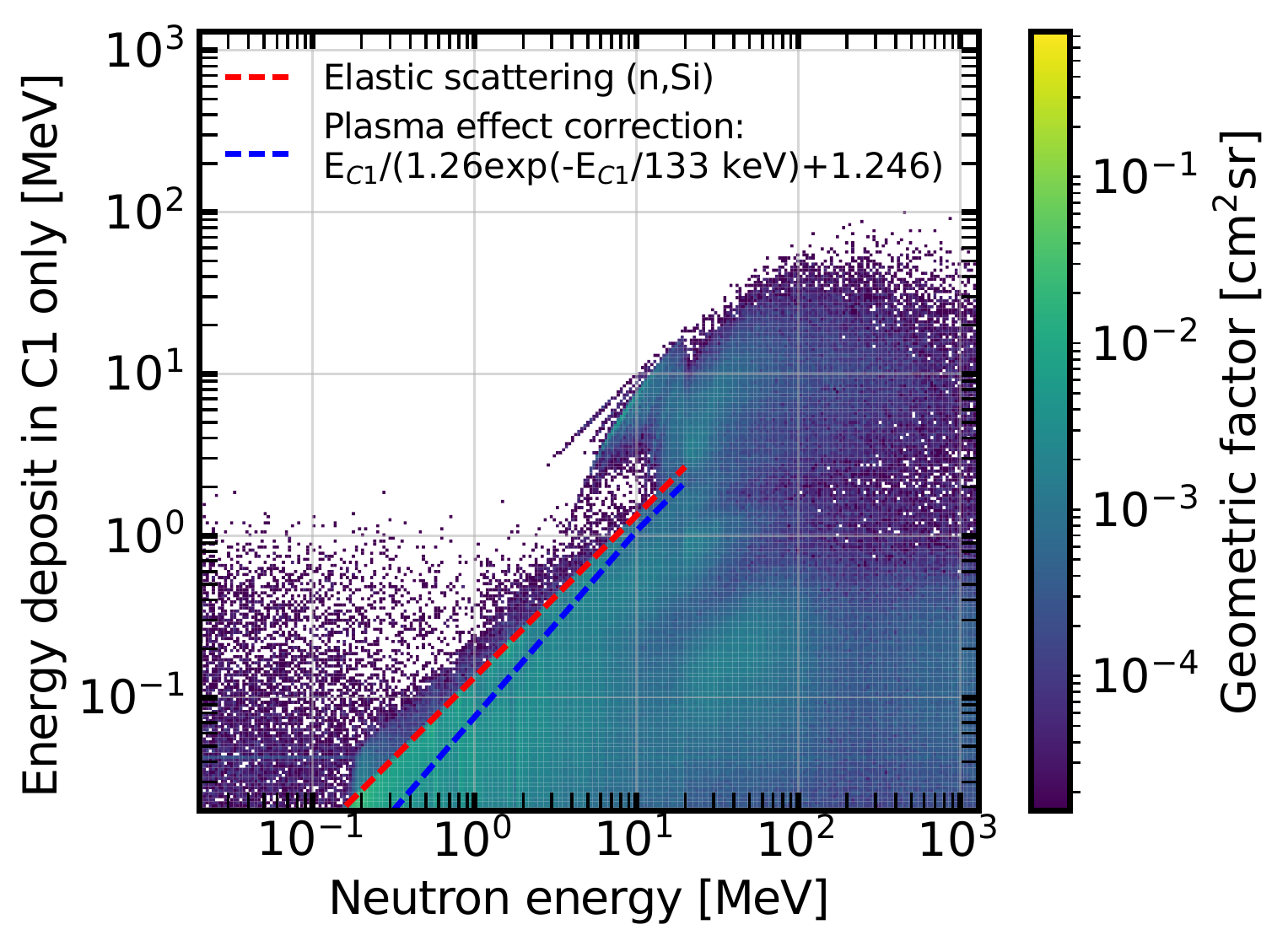}
  \caption{Geometry factor for the detection of neutrons. The red line shows the maximum recoil energy transfer. The blue dotted line shows the plasma effect.}
  \label{fig:neutron_geofac}
\end{subfigure}
\caption{Geometry factors for detecting gammas (left) and neutrons (right). The color scales are the same for both plots. See text for discussion. }
\label{fig:ng_geofacs}
\end{figure}

LND's top telescope consists of four segmented Si-\acs{SSD}s (A - D in Fig.~\ref{fig:LND-detector-stack}). The three lower detectors (B, C, and D) are packed as close together as possible using a special process developed by Canberra (now Mirion technologies), the provider of the \acs{SSD}s. The inner segment of the C detector, C1, is surrounded by the B and D detectors as well as the outer segment of the C detector, C2. Thus, any signal measured in C1 in anti-coincidence with B, D, and C2, will be due to a neutral particle (neutron or gamma ray). This configuration allows us to measure neutrons in the energy range $1 \lesssim E_n \lesssim 20$ MeV with a maximum energy deposition of $\sim 13\%$ of their primary energy (Fig.~\ref{fig:neutron_geofac}).
%
This figure shows the geometry factor for the detection of neutrons as a function of energy. In this plot, incident energy is shown along the $x$-axis, and the energy deposited in the detector is shown along the $y$-axis. The geometric factor for the measurement is color coded as shown in the color bar on the right and was simulated using GEANT4 version 10-1-patch-2\footnote{Using the physics lists QGSP\_BERT\_HP, G4EmLivermorePhysics, and G4RadioactiveDecayPhysics.}  \cite{agostinelli-etal-2003,allison-etal-2006,allison-etal-2016}. The dashed red line marks the previously discussed elastic scattering of neutrons with Si-nuclei, it marks the $13.3\%$ maximum energy transfer. Close inspection shows that the recoil nuclei deposit somewhat less energy than the allowed maximum which is due to the so-called plasma-effect \cite{tove-and-seibt-1967,moeller-2013}. Because the knock-on Si nuclei are highly ionizing and have only a very short range in Si, the electric field used to separate electron-hole pairs can not fully penetrate into the region of ionization and a fraction of the electrons and holes can recombine before they are separated. This effect results in a small correction which is indicated by the blue dotted line and is not included in the GEANT4 physics lists. 
%
Figure~\ref{fig:neutral_in_C1} shows the expected measured energy spectrum in C1 only (i.e.\,in anti-coincidence with the all other detectors and detector segments).

Continuing the discussion of Fig.~\ref{fig:ng_geofacs}, we now move to the left-hand panel, Fig.~\ref{fig:gamma_geofac}, which shows \acs{LND}'s response to $\gamma$-rays. It shows that $\gamma$-rays can be measured up to approximately 1 MeV. Above that energy, the geometry factor rapidly decreases. The diagonal line (highlighted with black dots) at the lowest energies is due to the photo-electric effect in which a $\gamma$-ray transfers energy to a photo-electron. This effect dominates \acs{LND}s $\gamma$-ray response at low energies because the recoil electrons do not have enough energy to exit the \acs{LND} detector and thus deposit all their energy in the 500 $\mu$m detector. To the right of the photo-electric effect, one sees the increasing importance of Compton scattering (highlighted with blue dots) which dominates for energies between $\sim$100 keV and a few MeV. At higher incident photon energies, the photo effect (shown with red dots) begins to contribute. \acs{LND}s detectors are virtually transparent for higher energies of $\gamma$ radiation. The horizontal lines which can be seen at low energy depositions are due to $\gamma$-rays emitted by various materials inside the \acs{LND} \acs{SH}, lead and tin in the electronics, and gadolinium from the Gd foils and absorber discussed in the following paragraphs. These lines are highlighted by colored dots and called out in the legend panel. The enhanced response at 17 MeV $< E_\gamma < $ 23 MeV (surrounded by a dashed black line) is due to photo-neutron reactions \cite{anderson-etal-1969}.

As has become obvious in the previous discussion, \acs{LND} can not truly discriminate between neutrons and $\gamma$-rays. Nevertheless, the different instrument response functions for $\gamma$-rays and neutrons (panels \ref{fig:gamma_geofac} and \ref{fig:neutron_geofac} in Fig.~\ref{fig:ng_geofacs}) result in different sensitivities to the two dominant neutral particle species. Energy depositions below 1 MeV are dominated by $\gamma$s whereas at higher deposited energies, neutrons gain in importance.

\subsection{Thermal Neutrons}

\begin{wrapfigure}[28]{l}{0mm}
\includegraphics[width=7cm]{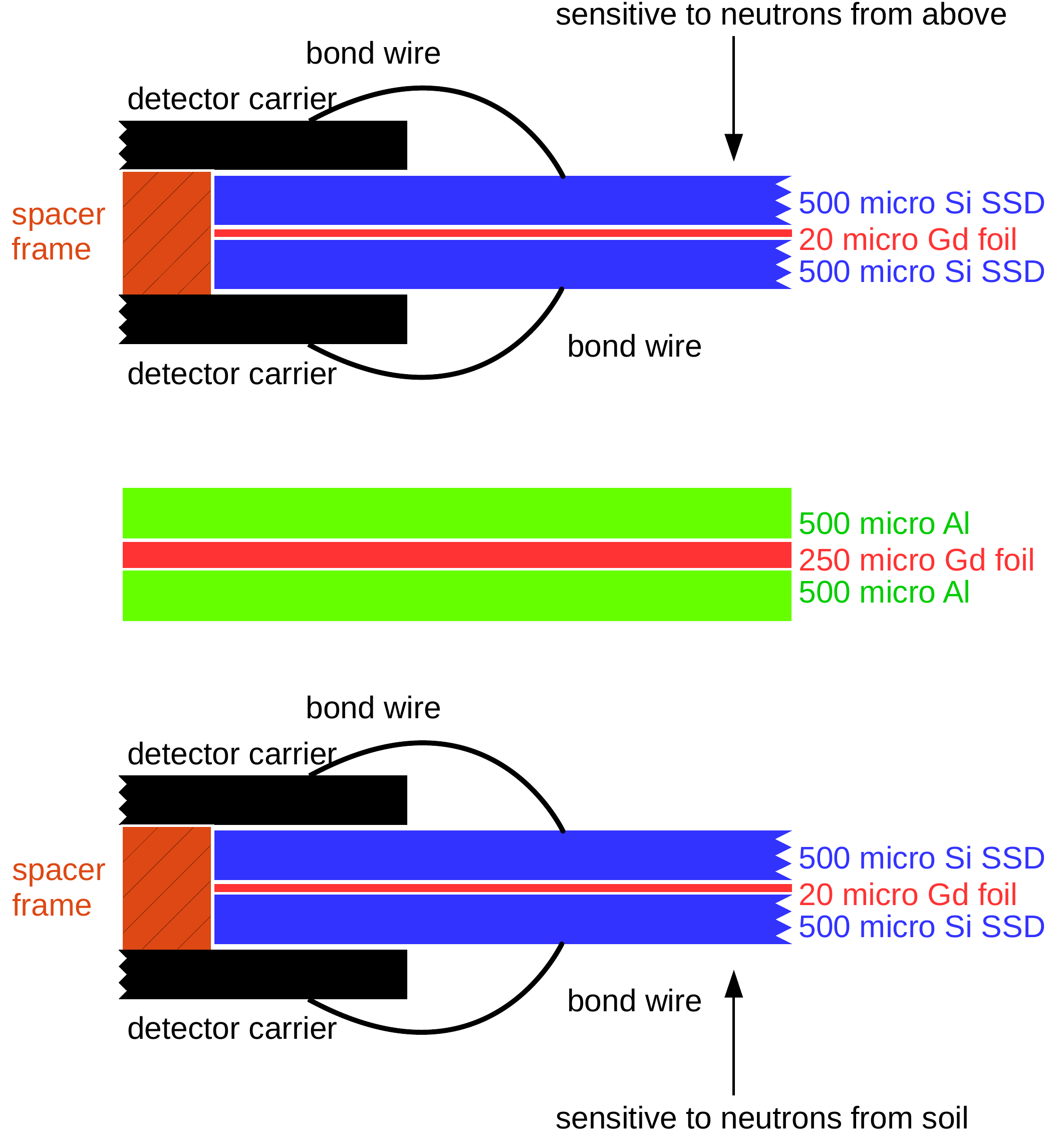}
\caption{Gd-sandwich detector concept. The very thin Gd foil captures thermal neutrons and emits conversion electrons. 
The thick Gd-foil encased in Al effectively shields the upper Gd-sandwich from thermal neutrons from the soil and the bottom Gd-sandwich from thermal neutrons from above.}
\label{fig:Gd-sandwich}
\end{wrapfigure}

We now move on to discuss the lower six detectors (E--J) of \acs{LND} which are mounted in two different sandwich configurations. In one, detectors E\&F and G\&H clamp a very thin ($\approx 20$ $\mu$m) Gd foil, as shown in red in Fig.~\ref{fig:Gd-sandwich}. The (natural) Gd-foil has a very large cross section\footnote{\cite{leinweber-etal-2006} and \cite{abdushukurov2010} give different values for the cross sections, here we cite those of \cite{leinweber-etal-2006}.} (48'800 barn) for thermal ($v_{\rm th} \sim 2200$m/s) neutron capture \cite{leinweber-etal-2006,abdushukurov2010} and therefore is used to detect thermal neutrons. The cross section is dominated by those of the $^{155}$Gd (60'700 barns) and $^{157}$Gd (254'000 barns) isotopes with natural abundances of 14.8\% and 15.7\% \cite{leinweber-etal-2006}. After a thermal neutron has been captured in a $^{155}$Gd or $^{157}$Gd nucleus, the excited $^{156}$Gd or $^{158}$Gd nucleus decays via internal conversion emitting a single conversion electron (in $\sim 80\%$ or $\sim 60\%$ of all cases) or $\gamma$-decay. Most of the conversion electrons have energies between $\sim 80$ and $\sim 300$ keV \cite{fogelberg-and-baecklin-1972}. The electrons have a range of several microns in the Gd foil and therefore can escape the foil and hit the neighboring sandwich detector.
Using the other \acs{LND} detectors as an anti-coincidence, this provides a clean measurement of thermal neutrons. The $\gamma$-ray from neutron capture can also be detected in anti-coincidence with surrounding detectors, providing an independent measurement of the thermal neutrons. The lowermost detector sandwich (I\&J) is a copy of the \acs{LND} B\&C-sandwich and serves as the final detector in the stack. Detector J serves as a discriminator for particles stopping in the particle telescope spanned by detectors A -- I.

Thermal neutrons emitted from the soil are very slow and a large fraction of them will return to the Moon because their speeds are less than the escape speed from the Moon, 2.4 km/s. To discriminate such ballistic thermal neutrons (which are sensitive to the subsurface proton (water) content) returning to the Moon from neutrons coming from beneath the lander, the E\&F sandwich is shielded from below by a $\sim 250$ $\mu$m thick Gd foil which is encased in two 500 $\mu$m thick Al sheets. The thickness of this Gd absorber was optimized to ensure that less than one in 1000 thermal neutrons can penetrate through the Gd. The Al encasing is thick enough that the electrons from the neutron capture cannot escape. Its thickness was also chosen to be the same as that of the Si-\acs{SSD}s to also absorb a similar fraction of the $\gamma$-rays from the neutron capture as in the Si-\acs{SSD}s. The G\&H sandwich then measures thermal neutrons from below and the E\&F sandwich measures thermal neutrons returning to the Moon from above. Because the return flux averages over a large area and the primary flux from below averages over a small area, the difference between the two gives us a measure of the local subsurface proton\footnote{Thermal neutrons are also good tracers of the fluorine content of the soil, measurements are thus somewhat ambiguous.} content. The detector geometry is summarized in Fig.~\ref{fig:Gd-sandwich}






\begin{figure}
  \centering
  \includegraphics[width=0.75\textwidth]{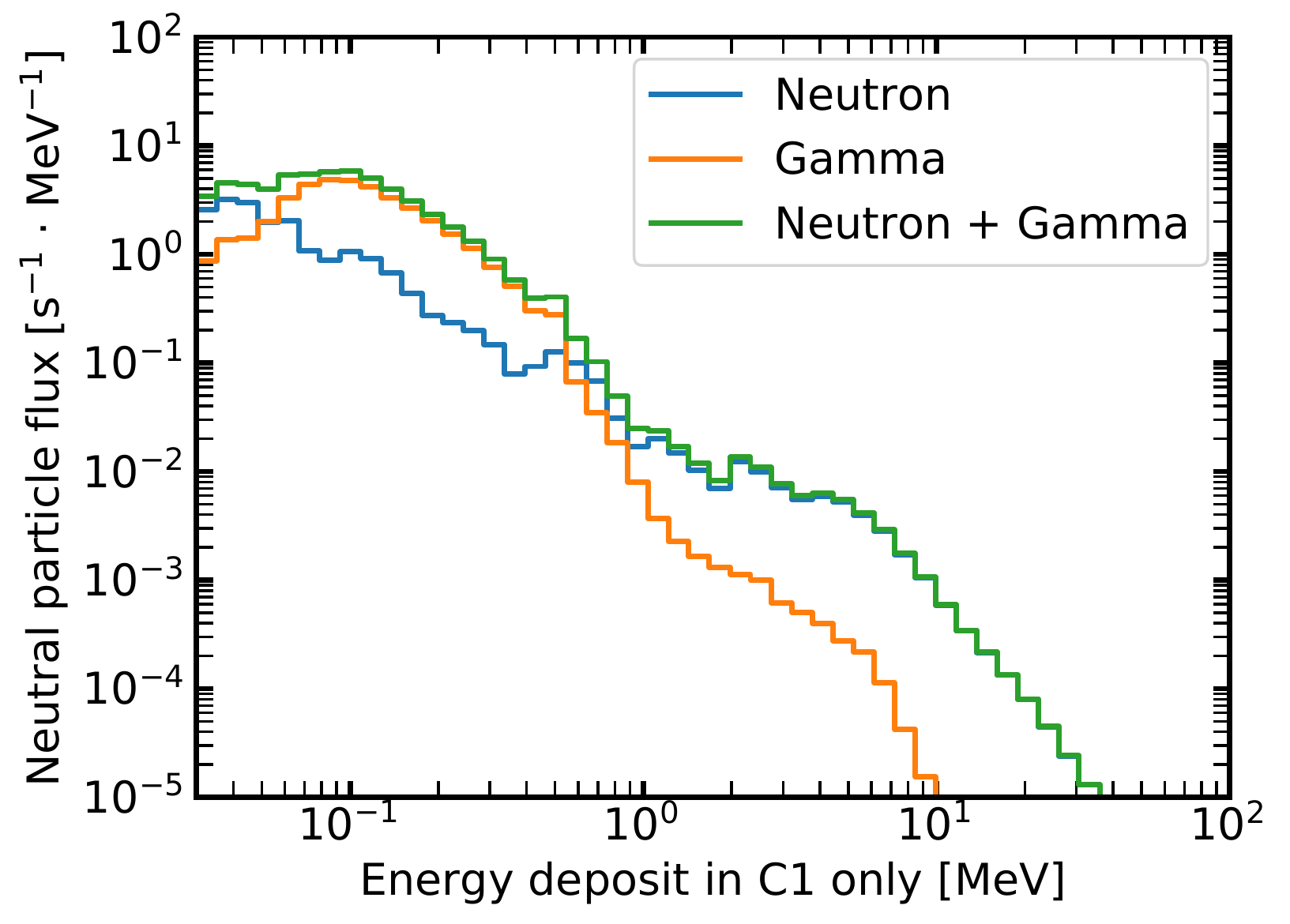}
  \caption{Modeled spectrum of energy deposits in C1 in anti-coincidence 
  with all other detector channels. Energy deposits above 1 MeV are dominated by neutrons.}
  \label{fig:neutral_in_C1}
\end{figure}


Model spectra of neutral particles are shown in Fig. \ref{fig:neutral_in_C1} and exhibit relatively low count rates. Integrated over all energies, we expect a count rate of about 1.3 counts per second in the absence of the RTGs/RHUs. At low energies, this count rate is dominated by gammas. As one can easily see in Fig.~\ref{fig:neutral_in_C1}, neutrons start to dominate the count rate above energy depositions of about 700 keV, at energies above $\sim 1$ MeV, we expect only about 0.006 gamma, but 0.06 neutron counts per second \footnote{Building up statistics over time, one may attempt to invert the neutron and gamma spectra similarly to what was done for the Radiation Assessment Detector (RAD) on NASA's Mars Science Laboratory (MSL)  \cite{koehler-etal-2011}, albeit with considerably lower counting statistics and fidelity of the result.}.

The Chang'E 4 lander and rover are designed to survive many harsh lunar day and night cycles. Therefore, the lander carries one Radioisotope Thermoelectric Generator (RTG) for power and three Radioisotope Heater Units (RHUs). The background from the RTG and RHUs was measured at the 18-th Research Institute, China Electronics Technology Group Corporation in Tianjin, China. The results are reported in \cite{hou-etal-2019}. 







\section{Data Products}
\label{sec:data-products}

The measurement requirements given in section~\ref{sec:measurement-requirements} map to the \acs{LND} data products which are summarized in Tab.~\ref{tab:data}. They are a mix of some data products at high time resolution (one minute), medium time resolution (10 minutes), and low time resolution data (one hour). 
The different cadences of \acs{LND}'s data products allow the investigation of rapid variations such as are, e.\,g., expected during the onset of a solar particle event. At the same time, no telemetry is wasted with sending {\em all} information at this high cadence. 
This chapter gives a detailed description of LND's data processing and data products.

\subsection{Data Processing Inside \acs{LND}}
\label{sec:data-processing}

\begin{sidewaysfigure}
  \centering
  \vspace{13cm}
\includegraphics[width=0.9\textwidth]{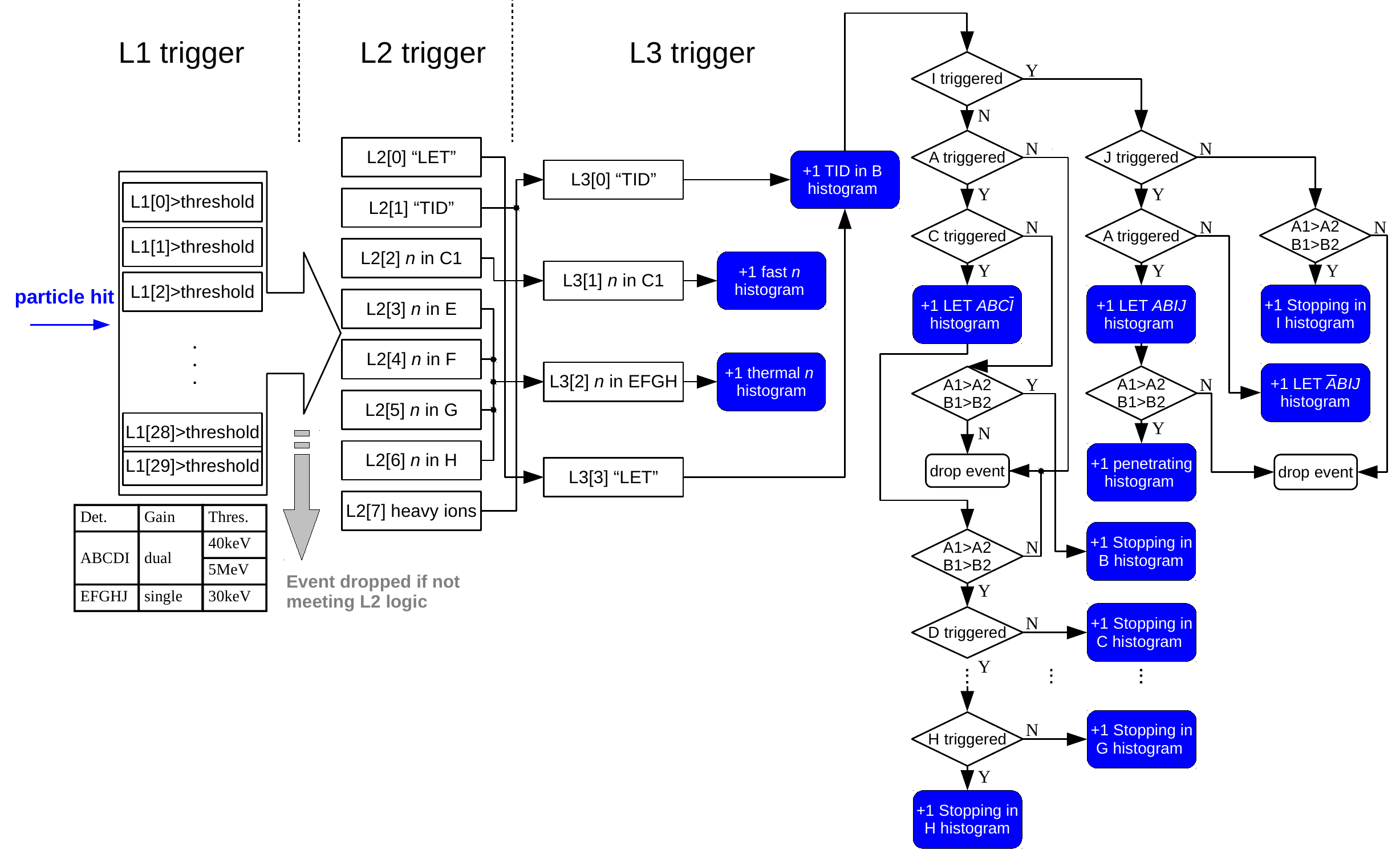}
  \caption{LND's trigger logic is divided into three levels (L1, L2, and L3). Processing is distributed into the analog FPGA (L1 \& L2) and digital FPGA (L3).}
  \label{fig:L1L2L3_diagram}
\end{sidewaysfigure}

LND data processing is summarized in Fig.~\ref{fig:L1L2L3_diagram} which we discuss from left to right here. When a particle hits one or more of LND's detector segments, the L1 trigger checks which of LND's 30 detector (gain) channels registered an energy deposition larger than their trigger thresholds. These are given in the box beneath the left-most entry of Fig.~\ref{fig:L1L2L3_diagram}. This information as well as the measured values are passed on to the L2 trigger which performs a number of tasks. It checks whether the event has triggered detectors which are relevant for the LET data product (different detectors can contribute). If so, the counter for valid LET events (L2$[0]$) is incremented by one. Similarly, if an event is classified as fulfilling the conditions to contribute to the determination of TID, the counter for valid TID events (L2$[1]$) is incremented by one. This is repeated for neutral particles detected in C1 (L2$[2]$) and detectors E -- F (L2$[3]$ -- L2$[6]$). Such events are valid if energy was deposited in only one of these detectors, i.e., in anti-coincidence with all other detectors. Finally, if an event is identified as being due to a heavy ion, the L2$[7]$ counter is incremented by one. An event which does not satisfy these trigger conditions is dropped. All processing up to this point is performed in the analog FPGA (see Fig.~\ref{fig:block_diagram}).

After this processing by the L2 trigger, a valid event is passed on to the L3 trigger which is implemented in the digital FPGA. This trigger further classifies events and increments the appropriate counters. We being this discussion with the L3[1] and L3[2] triggers because they are the simplest ones. An event which was identified by the L2 trigger as an event seen only in C1 (in anti-coincidence with all other detector segments) is further analysed and the spectrum of energy depositions in the C1 detector is incremented accordingly. This is indicated by "$+1$ fast n histogram" in Fig.~\ref{fig:L1L2L3_diagram}. The same is done for events seen in detectors E -- H only (L3[2] trigger: "$+1$ thermal n histogram").  Because these events only triggered single detectors, their processing ends here. For all other events, a more complicated processing chain is applied. An event which satisfies the Ls[0] condition to contribute to TID has triggered LND's B detector. Depending on the energy deposited in detector B, the corresponding histogram counter is incremented by one. While TID is only a number, LND can not measure it as such. The spectrum of energy depositions thus accumulated needs to be summed on Earth to derive TID, see Sec.~\ref{sec:dosi}. This event is further processed by checking whether it also triggered detector I, as depicted in Fig.~\ref{fig:L1L2L3_diagram}. Events which are classified as L3[3] events have triggered the B detector and at least one other detector, they are processed in the same manner. The flow chart in the right of Fig.~\ref{fig:L1L2L3_diagram} describes the further processing chain.


The histograms which are thus accumulated by LND can be one-dimensional or two-dimensional. TID and some LET spectra are accumulated in one-dimensional histograms whereas events that triggered several detectors augment two-dimensional histograms. We refer to the collection of these histograms as the Xmas plot (see Sec.\ref{sec:xmas} and Fig.~\ref{fig:xmas2}). The quantities used to calculate the entry "pixel" in the Xmas plot are calculated according to the quantities summarized in Tab.~\ref{tab:axis-defs}, the corresponding energy ranges are given in Tab.~\ref{tab:stop-energies}. The individual entry types are discussed in the following subsections.






\begin{table}
  \centering
  \begin{tabular}{|lcp{7cm}|}\hline\hline
    \multicolumn{3}{|c|}{\underline{\bf 1-minute data products}}\\
    {\bf Data product} & {\bf \# counters} & {\bf Comments} \\\hline
    Protons          & {14}  & 2 counters for B, one each in C, D, E, F, G, H, and I, {5} for penetrating. See Tab.~\ref{tab:stop-energies} for energy ranges. \\
    Electrons$^a$        & 7  & 4 counters for electrons stopping in B, C, D, 3 counters for electrons stopping in E and F, one for G, H, and I combined. See Tab.~\ref{tab:e-electrons} for energy ranges.\\
    Neutrals         & 4  & 4 counters for $20 < E < 200$, $200 < E < 10^3$, $10^3 < E < 10^4$, $10^4 < E < 10^5$ keV \\
    Dosimetry$^b$    & 8  & 8 counters for $20 < E < 95$, $95 < E < 320$, $320 < E < 1076$~keV, and $1.076 < E < 3.62$, $3.62 < E < 12.2$, $12.2 < E < 41$, $41 <  E < 138$, $138 < E < 390$~MeV\\
    LET spectrum$^b$ & 8  & same bins as dosimetry \\\hline\hline

    \multicolumn{3}{|c|}{\underline{\bf 10-minute data products}}\\
     {\bf Data product} & {\bf \# counters} & {\bf Comments} \\\hline
    Thermal neutrons & 8  & 1 {each} for E1, E2, F1, F2, G1, G2, H1, H2, covering $\sim 55 < E < \sim 75$~keV\\
    Electrons$^c$        & 16 & 16 counters for electrons stopping in E, F, G, H {(4 counters for each detector.)} See Tab.~\ref{tab:e-electrons} for energy ranges.\\
    $^3$He           & 5  & 3 counters in B, one each in C and D (stopping). See Tab.~\ref{tab:stop-energies} for energy ranges. \\
    $^4$He           & 10 & 3 counters in B, one box in C, D, E, F, G, H, and I. See Tab.~\ref{tab:stop-energies} for energy ranges.  \\
    CNO              & 9  & 2 counters in B, one box in C, D, E, F, G, H, and I. See Tab.~\ref{tab:stop-energies} for energy ranges.  \\
    heavy ions       & 9  & 2 counters in B, one box in C, D, E, F, G, H, and I. See Tab.~\ref{tab:stop-energies} for energy ranges.  \\
    penetrating      & {7} & $^3$He, $^4$He, and heavy ions. See Tab.~\ref{tab:stop-energies} for energy ranges.  \\\hline\hline

    \multicolumn{3}{|c|}{\underline{\bf 60-minute data products}}\\
                {\bf Data product} & {\bf \# entries} & {\bf Comments} \\\hline
                ``Xmas plot'' & $150 \cdot 64$ & Full information \\
                PHA words     & n/a        & 16 PHA buffers of 1 kByte each per hour for diagnostic and calibration purposes.\\\hline\hline
  \end{tabular}
  \caption{List of \acs{LND} science data products which are included in telemetry at 1, 10, and 60-minute time resolution as indicated by the headings. $^a$Defined in Tab.~\ref{tab:e-electrons}. $^b$ Bin edges will be adjusted during the course of the mission. $^c$ The counters are defined in Tab.~\ref{tab:e-electrons}.}

\label{tab:data}
\end{table}

\begin{figure}
  \centering
  \hspace{-1.cm}
  \includegraphics[height=0.96\textwidth,angle=90]{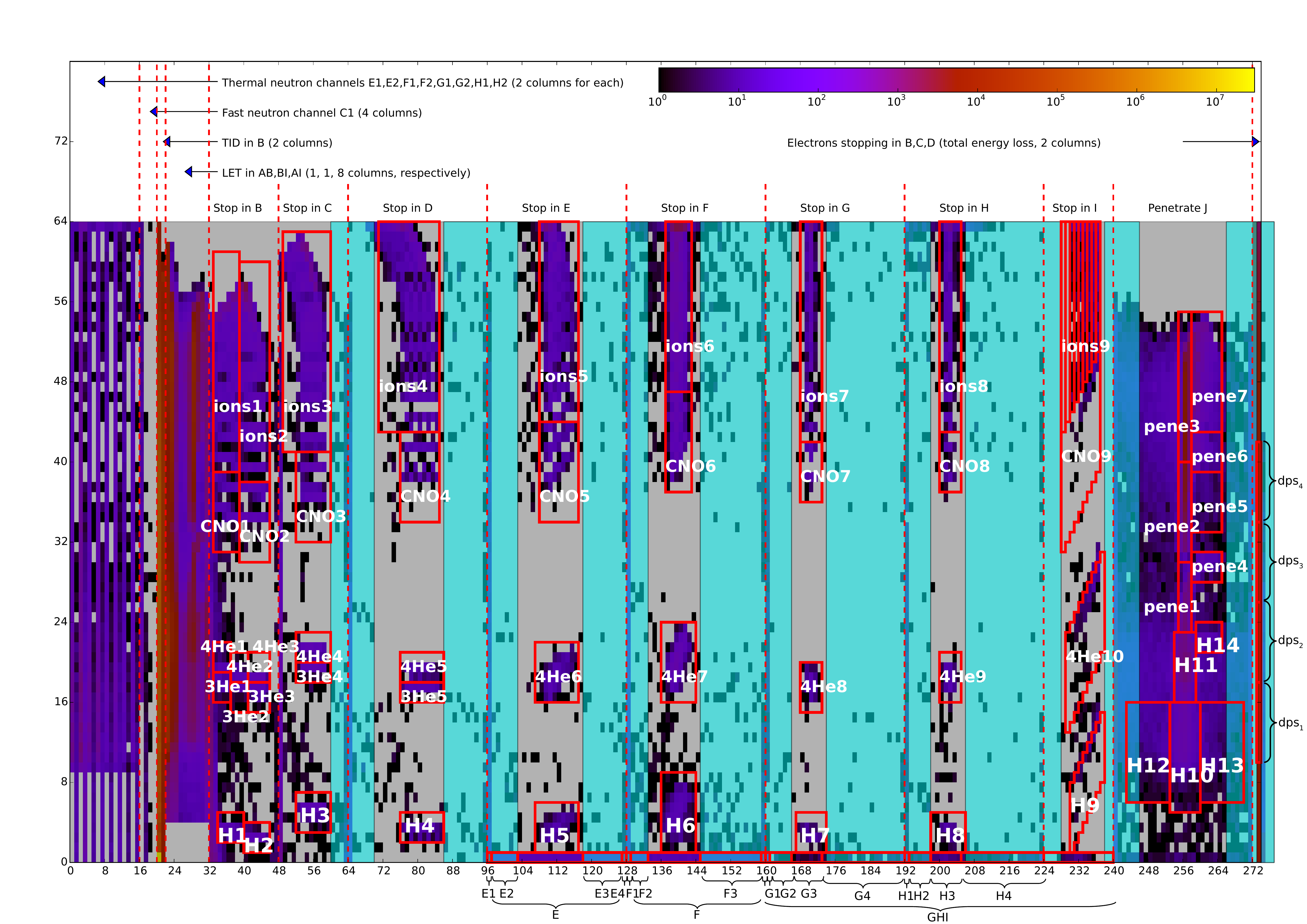}  \caption{LND ``Xmas plot''. It is divided into several vertical regions which contain information about different particle populations. The principal divisions are indicated in this figure which is based on GEANT4 simulations. Heavy ions are over-represented to make them visible. The regions indicated with red boundaries in this Figure will be adapted to reflect updated calibration values. They are labeled according to the entries given in Tabs.~\ref{tab:xmas1}, \ref{tab:stop-energies}, and \ref{tab:e-electrons}. The regions shaded in cyan are not transmitted back to Earth.}
  \label{fig:xmas2}
\end{figure}

\subsection{Figure \ref{fig:xmas2} -- the ``Xmas Plot''}
\label{sec:xmas}

Fig.~\ref{fig:xmas2} shows a "Xmas plot\footnote{We call it the ``Xmas plot'' because we sent this data summary plot to colleagues around the world as our Xmas card in 2017. We had to choose a name for this plot.}" which is LND's principal data product, all others are derived from it. It contains all data of the particles measured by LND. The Xmas plot is actually an image of LND's accumulator memory, the different regions are explained in more detail in Tab.~\ref{tab:xmas1} and are summarized below. Each pixel of the Xmas plot is a counter which is incremented as discussed below. LND memory contains two Xmas plots, an active and an inactive one. Upon measurement of a particle, the active one is updated as described below. This accumulation is ended after 3600 seconds when the Xmas plot is turned to inactive mode. At this moment, the counters in the previously inactive Xmas plot are reset to zero and this memory region (Xmas plot) is turned to active mode. The previously active Xmas plot is now inactive and is red out and sent to telemetry over the course of the next hour. Thus the two Xmas plots alternate between active and inactive mode and provide continuous data coverage. Regions shaded in cyan in Fig.~\ref{fig:xmas2} are not put into telemetry because they don't contain physical data and are dominated by noise. 

The Xmas plot is divided up column wise into different regions which are defined in Tab.~\ref{tab:xmas1}. Each column has 64 rows. Each point in this $274\times 64$ matrix serves as a counter which is incremented when the LND L3 trigger identifies this to be the correct entry. The integer counters are 4 Bytes, 6 of the 32 bits serve as Hamming-encoding single-even upset correction.

We now proceed to give the explanation of the Xmas plot starting from the left and moving through it to the right. The left most region of the Xmas plot contains the accumulators for neutral particles (Sec.~\ref{sec:neutrals}), it is followed (from left to right) by two regions for the dosimetric quantities total ionizing dose (TID) and linear energy transfers (LETs) (Sec.~\ref{sec:dosi}). To the right of these are memory regions for the accumulation of particles stopping in LNDs detectors B--I (Sec.~\ref{sec:charged}). Penetrating particles are mapped to the region indicated by ``Penetrate J'' (Sec.~\ref{sec:penetrating}). The single column on the very right contains electron data, as does the narrow strip at the bottom of the plot (Sec.~\ref{sec:electrons}). The summed counts in the regions indicated by red counters and white names are read out and telemetered at a cadence of 1 or 10 minutes, while the entire Xmas plot with the full information is put into telemetry every hour. The 10 and 1 minute data are explained in more detail in sections~\ref{sec:10-min} and ~\ref{sec:1-min}.

Fig.~\ref{fig:xmas2} shows an annotated Xmas plot, the annotations are referred to in the following paragraphs.

\begin{sidewaystable}
  \centering
  \vspace{12cm}
  \begin{tabular}{|lrr|l|}\hline\hline
     {\phantom{bla}}                  & \multicolumn{2}{c|}{\bf Columns} & \\
    {\bf Xmas Plot Region} & {\bf start} & {\bf end} & \\\hline
    {\bf Neutral particles} & {\bf 0} & {\bf 19} & \\\hline
    detected in E1              & 0 & 1   & $E1 \cdot \overline{E2} \cdot \overline{A} \cdot \overline{B} \cdot \overline{C} \cdot \overline{D} \cdot \overline{F} \cdot \overline{G} \cdot \overline{H} \cdot \overline{I} \cdot \overline{J}$\\
    detected in E2              & 2 & 3   & $E2 \cdot \overline{E1} \cdot \overline{A} \cdot \overline{B} \cdot \overline{C} \cdot \overline{D} \cdot \overline{F} \cdot \overline{G} \cdot \overline{H} \cdot \overline{I} \cdot \overline{J}$\\
    detected in F1              & 4 & 5   & $F1 \cdot \overline{F2} \cdot \overline{A} \cdot \overline{B} \cdot \overline{C} \cdot \overline{D} \cdot \overline{E} \cdot \overline{G} \cdot \overline{H} \cdot \overline{I} \cdot \overline{J}$\\
    detected in F2              & 6 & 7   & $F2 \cdot \overline{F1} \cdot \overline{A} \cdot \overline{B} \cdot \overline{C} \cdot \overline{D} \cdot \overline{E} \cdot \overline{G} \cdot \overline{H} \cdot \overline{I} \cdot \overline{J}$\\
    detected in G1              & 8 & 9   & $G1 \cdot \overline{G2} \cdot \overline{A} \cdot \overline{B} \cdot \overline{C} \cdot \overline{D} \cdot \overline{E} \cdot \overline{F} \cdot \overline{H} \cdot \overline{I} \cdot \overline{J}$\\
    detected in G2              & 10 & 11 & $G2 \cdot \overline{G1} \cdot \overline{A} \cdot \overline{B} \cdot \overline{C} \cdot \overline{D} \cdot \overline{E} \cdot \overline{F} \cdot \overline{H} \cdot \overline{I} \cdot \overline{J}$\\
    detected in H1              & 12 & 13 & $H1 \cdot \overline{H2} \cdot \overline{A} \cdot \overline{B} \cdot \overline{C} \cdot \overline{D} \cdot \overline{E} \cdot \overline{F} \cdot \overline{G} \cdot \overline{I} \cdot \overline{J}$\\
    detected in H2              & 14 & 15 & $H2 \cdot \overline{H1} \cdot \overline{A} \cdot \overline{B} \cdot \overline{C} \cdot \overline{D} \cdot \overline{E} \cdot \overline{F} \cdot \overline{G} \cdot \overline{I} \cdot \overline{J}$\\
    detected in C1     & 16 & 19  & $C1 \cdot \overline{C2} \cdot \overline{A} \cdot \overline{B} \cdot \overline{D} \cdot \overline{E} \cdot \overline{F} \cdot \overline{G} \cdot \overline{H} \cdot \overline{I} \cdot \overline{J}$\\\hline
    {\bf Dosimetry}         & {\bf 20} & {\bf 31}  & \\\hline
    TID in B          & 20 & 21  & $B$ (if energy deposit in B is bigger than 40keV)\\
    LET $ABC\overline{I}$      & 22 & 22  & $A \cdot B \cdot C \cdot \overline{I}$\\
    LET $\overline{A}BIJ$         & 23 & 23  & $\overline{A} \cdot B \cdot I \cdot J$\\
    LET $ABIJ$         & 24 & 31  & $A \cdot B \cdot I \cdot J$\\\hline
    {\bf Charged particles} & {\bf 32} & {\bf 273}  & \\\hline
    stopping in B   & 32  & 47   & $A \cdot B \cdot \overline{C} \cdot \overline{I} \cdot (E_{A1}>E_{A2}) \cdot (E_{B1}>E_{B2})$\\
    stopping in C   & 48  & 59   & $A \cdot B \cdot C \cdot \overline{D} \cdot \overline{I} \cdot (E_{A1}>E_{A2}) \cdot (E_{B1}>E_{B2})$ \\
    stopping in D   & 70  & 85   & $A \cdot B \cdot C \cdot D \cdot \overline{E} \cdot \overline{I}\cdot (E_{A1}>E_{A2}) \cdot (E_{B1}>E_{B2}) \cdot (E_{D1}>E_{D2})$ \\  
    stopping in E   & 103 & 117  & $A \cdot B \cdot C \cdot D \cdot E \cdot \overline{F} \cdot \overline{I}\cdot (E_{A1}>E_{A2}) \cdot (E_{B1}>E_{B2}) \cdot (E_{D1}>E_{D2})$ \\
    stopping in F   & 133 & 144  & $A \cdot B \cdot C \cdot D \cdot E \cdot F \cdot \overline{G} \cdot \overline{I}\cdot (E_{A1}>E_{A2}) \cdot (E_{B1}>E_{B2}) \cdot (E_{D1}>E_{D2})$\\
    stopping in G   & 166 & 173  & $A \cdot B \cdot C \cdot D \cdot E \cdot F \cdot G \cdot \overline{H} \cdot \overline{I}\cdot (E_{A1}>E_{A2}) \cdot (E_{B1}>E_{B2}) \cdot (E_{D1}>E_{D2})$\\
    stopping in H   & 198 & 205  & $A \cdot B \cdot C \cdot D \cdot E \cdot F \cdot G \cdot H \cdot \overline{I}\cdot (E_{A1}>E_{A2}) \cdot (E_{B1}>E_{B2}) \cdot (E_{D1}>E_{D2})$\\
    stopping in I   & 228 & 237  & $A \cdot B \cdot C \cdot D \cdot I \cdot \overline{J}\cdot (E_{A1}>E_{A2}) \cdot (E_{B1}>E_{B2}) \cdot (E_{D1}>E_{D2})$\\
    Penetrating     & 246 & 265  & $A \cdot B \cdot I \cdot J \cdot (E_{A1}>E_{A2}) \cdot (E_{B1}>E_{B2}) $\\ \hline
    Electrons              & 273 & 273  & stopping in B if $16\cdot \log_2((E_A+E_{B1})\cdot E_A/(8000 \cdot 4000))<0$\\
                           &     &      & stopping in C if $16\cdot \log_2((E_A+E_{B1}+E_C)\cdot E_A/(10000 \cdot 4000))<-16$\\ 
                           &     &      & stopping in D if $16\cdot \log_2((E_A+E_B+E_C+E_{D1})\cdot E_A/(1200 \cdot 6000))<0$\\    \hline\hline
   \end{tabular}
  \caption{Column ranges for the different regions in the Xmas plot. Each column contains 64 rows or entries. Columns which are not mentioned in this table are not transmitted back to Earth and are shaded in Fig.~\ref{fig:xmas2}. }
  \label{tab:xmas1}
\end{sidewaystable}


\subsection{Neutral Particles}
\label{sec:neutrals}

Neutral particle measurements are based on measurements in single-detectors or detector segments in anti-coincidence with all other detectors/segments. The energy deposits measured in individual detectors are histogrammed and telemetered down as part of the Xmas plot, some lower-resolution data products are sent at higher time resolution, as summarized in Tab.~\ref{tab:xmas1}. Because the Chang'E 4 lander has an RTG and three RHUs, small energy deposits in the single detector are dominated by the background from the lander's RTG and RHUs. This needs to be accounted for when interpreting \acs{LND}'s measurements and using them to predict human exposure on the surface of the Moon.

\subsubsection{Thermal Neutrons}
\begin{figure}
  \centering
  \includegraphics[width=0.98\textwidth,clip=]{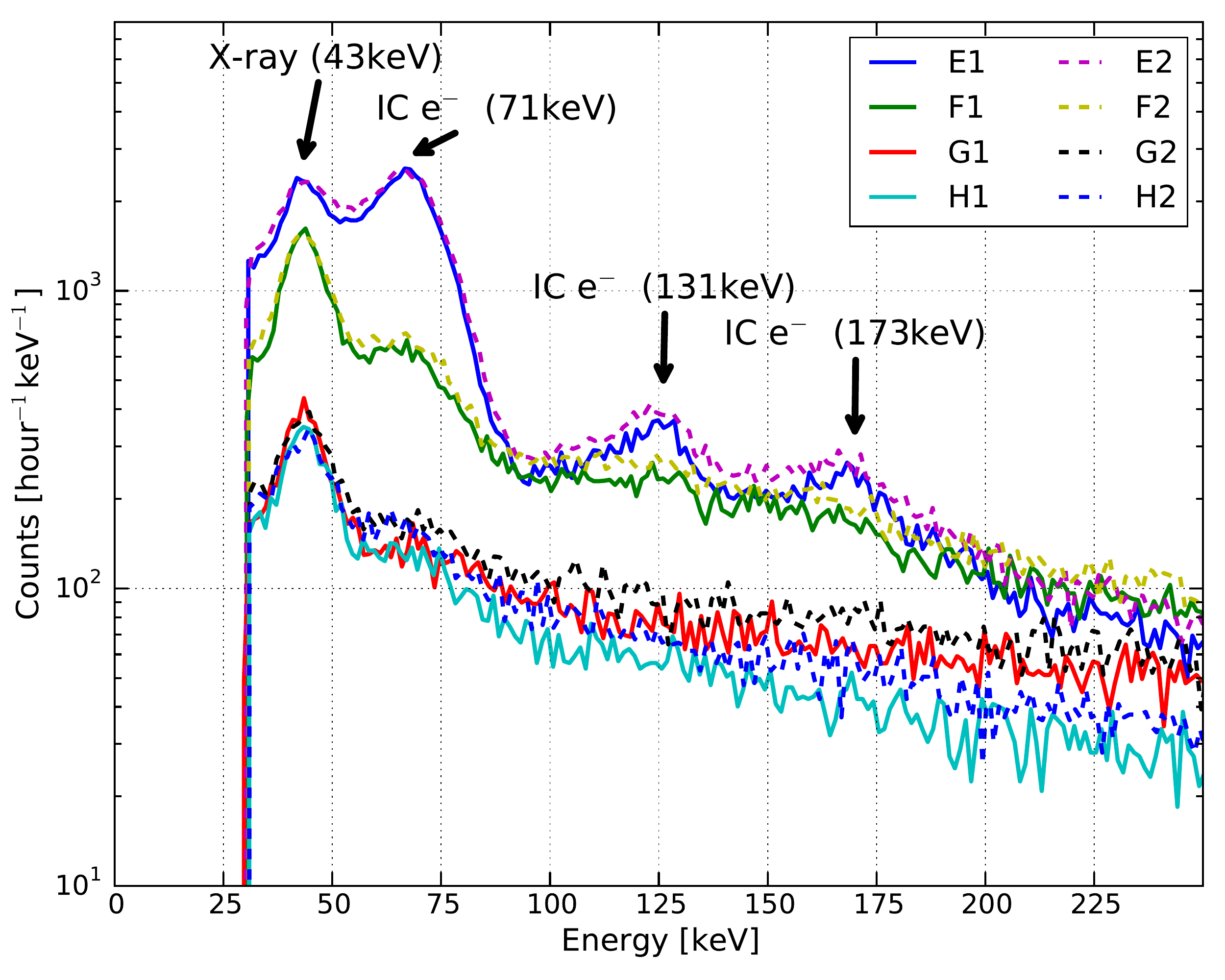}
  \caption{Thermal neutron measurements acquired with the LND FS at the Atominstitut (\acs{ATI}) in Vienna in September 2017 during one hour. Solid and dashed lines show measurements in the inner and outer segments (1 and 2) of LND's detectors E, F, G, and H. The beam hit LND from the front, so detectors G and H do not see thermal neutrons. Signatures include X-rays at 43 keV and internal conversion electrons at 71, 131, and 173 keV.}
  \label{fig:E1-histo}
\end{figure}

The left-most 16 columns contain the histograms of the energy deposits in detectors E, F, G, and H. Each detector has two segments (1: central, 2: outer segment), so there are 8 histograms in total. The energy deposit is accumulated in 128 channels, it is wrapped around once to fit into the format of the Xmas plot, thus each histogram covers two columns in the Xmas plot. An example will help understand the structure:

The histogram of energies measured in E1, $E_{E1}$, is shown in the two left-most columns of Fig.~\ref{fig:xmas2} (see also Tab.~\ref{tab:xmas1}). The 128 bins are accumulated based on the following formula:
\begin{equation}
  {\rm bin \#} = 16\cdot \log_2(E_{E1}/20),
  \label{eq:E1}
\end{equation}
where the energy values $E_{E1}$ and 20 are in keV. Such histograms are shown in Fig.~\ref{fig:E1-histo} and are shown in a color scale in the Xmas plot. Thus, to extract the E1 histogram one only needs to extract the two left-most columns of the Xmas plot. This is fully analogous for all thermal neutron detectors. 

The number of counts due to thermal neutrons is determined by taking the differences $E-F$ and $H-G$. We only expect a small fraction of thermal neutrons to be measured in F and in G, thus these two detectors can serve as background measurements but at the cost of a somewhat lower efficiency for detecting thermal neutrons. The number of thermal neutron counts is then divided by the detection efficiency for thermal neutrons, $\eta_{tn}\approx 1/50$, which was established during the calibration at \acs{ATI} in Vienna in September 2017. Together with the area of \acs{LND}'s detectors (given in Fig.~\ref{fig:detector}) \acs{LND} can thus determine the flux of thermal neutrons on the surface of the Moon.

\subsubsection{Fast Neutrons (and Gamma Radiation)}

The inner segment of the C detector, C1, is also read out in anti-coincidence with all other \acs{LND} detectors. It thus responds nearly exclusively to neutral particles which do not interact with the surrounding detectors and the outer segment of C, C2. Its energy histogram is stored in 256 bins and is wrapped around three times in the Xmas plot, see Fig.~\ref{fig:xmas2}.  The histogram bins are calculated according to eq.~\ref{eq:C1} 
\begin{equation}
  {\rm bin \#} = 16\cdot \log_2(E_{C1}/20),
  \label{eq:C1}
\end{equation}
where $E_{C1}$ and 20 are in keV, i.e., exactly as in eq.~\ref{eq:E1}. Thus the count rate of neutral particles is easily computed.

Figure~\ref{fig:ng_geofacs} shows the geometry factors $G_\gamma$ for $\gamma$-rays (Fig.~\ref{fig:gamma_geofac}) and $G_n$ neutrons (Fig.~\ref{fig:neutron_geofac}) in C1. The vector of differential count rates in the energy bins of C1, $\vec{z}$, is then given by
\begin{equation}
\vec{z} = G_\gamma \cdot \vec{J}_\gamma + G_n \cdot \vec{J}_n
    \label{eq:CR-ng}
\end{equation}
where $\vec{J}_\gamma$ and $\vec{J}_n$ are the vectors containing the differential fluxes of $\gamma$-rays and neutrons, respectively. Inspection of Fig.~\ref{fig:ng_geofacs} shows that an energy deposition in C1 above $\sim$ 1 MeV is very unlikely to be due to $\gamma$-rays, but much more probable for neutrons with energies $E_n > 10$ MeV, as is dicussed in more detail in Sec.~\ref{sec:neutral}. For energy depositions below $\sim 1$ MeV, \acs{LND} can not discriminate between neutrons and $\gamma$-rays. Inferences about the neutron and $\gamma$ spectra will have to rely on an extrapolation of the higher-energy neutron spectrum which should be based on modeling results. The process to obtain fast neutron spectra is based on similar work already performed for \acs{MSL}/\acs{RAD} \cite{koehler-etal-2014,guo-etal-2017}.

\subsection{Dosimetry: Total Ionizing Dose, \acs{TID}, and Linear Energy Transfer, \acs{LET}}
\label{sec:dosi}

\subsubsection{Total Ionizing Dose}

While Total Ionizing Dose (\acs{TID}) is only one number, it can not be computed by \acs{LND} and therefore needs to be calculated on ground. To measure TID the energy spectrum of all particles measured in detector B (i.e., in {\em both} the inner {\em and} the outer segments of B, B1 and B2) is accumulated in 128 bins according to eq.~\ref{eq:TID}.
\begin{equation}
  {\rm bin \#} = 8\cdot \log_2(E_B/20),
  \label{eq:TID}
\end{equation}
where $E_B$ and 20 are measured in keV. On ground, \acs{TID} is then computed using eq.~\ref{eq:TID2},
\begin{equation}
  {\rm TID} = \sum_{i=8}^{127} \sqrt{E_i\cdot E_{i+1}}\ \cdot\ {\rm counts}_i,
  \label{eq:TID2}
\end{equation}
where $i$ is the bin number and 
\begin{equation}
  E_i = 20 \cdot 2^{i/8}\ \mbox{[keV]}
  \label{eq:TID3}
\end{equation}
Note that the average energy of the bin is expressed here as the harmonic mean. \acs{LND} uses logarithmically spaced bins. It could also be expressed by the usual arithmetic mean, or also by the logarithmic mean which is larger than the harmonic mean and smaller than the arithmetic mean. Which of these expressions is used for the average energy of a bin is ambiguous. The choice should depend on the observed spectrum of energy deposits in B. Values reported in \cite{zhang-etal-2019} are calculated according to eq.~\ref{eq:TID3}.

\begin{figure}
   \centering
  \begin{subfigure}{.475\textwidth}
  \centering
  \includegraphics[width=1.0\textwidth,clip=]{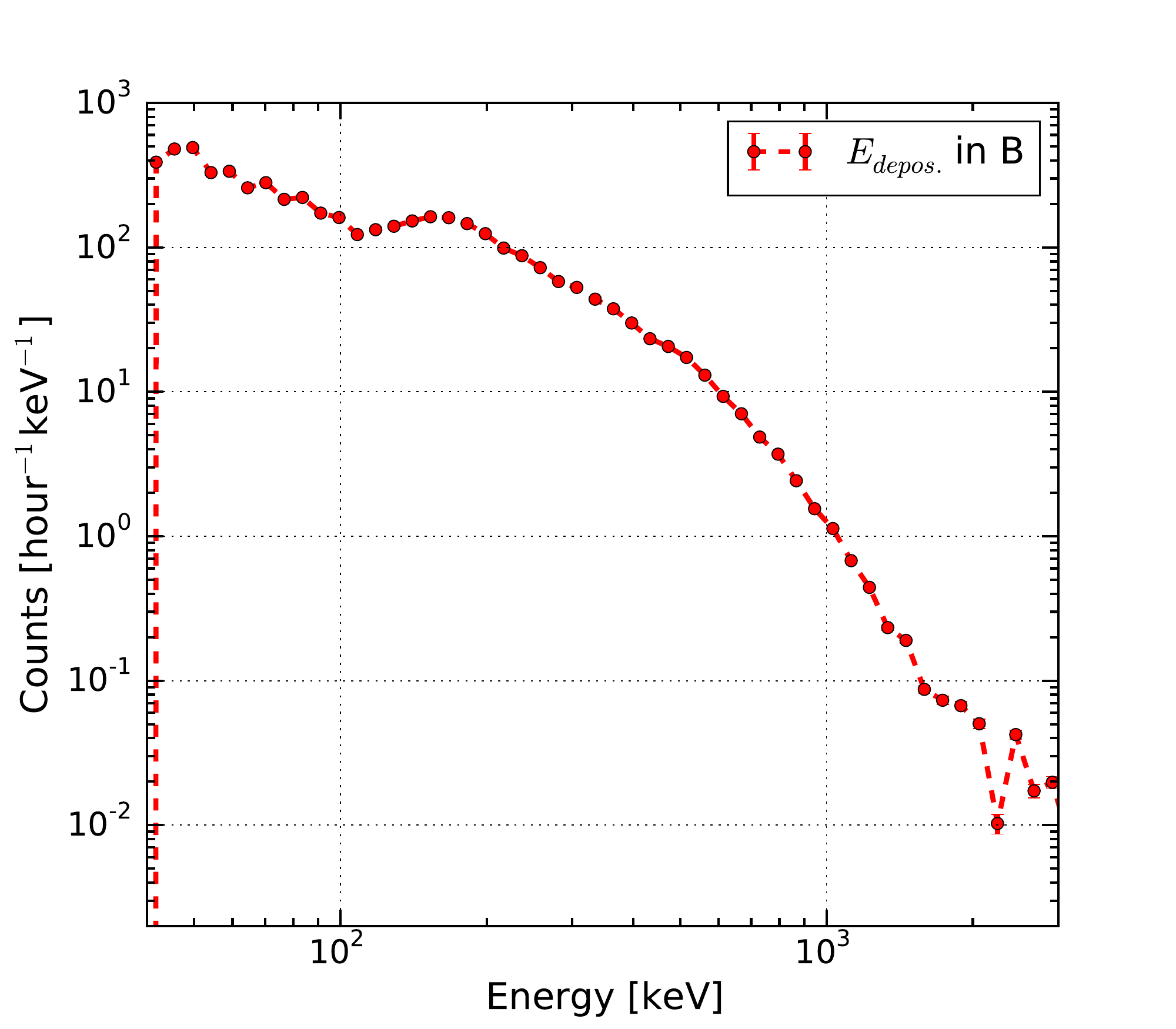}
  \caption{Spectrum of energy depositions in the B detector when \acs{LND} was standing upright, pointing at the zenith in our laboratory in Kiel. The threshold of $\sim 40$ keV is clearly seen, as are the cosmic muons between 100 and 200 keV. The blue curve was acquired using \acs{LND}s high-speed streaming mode, the red dots are data as stored in the ``X-mas card''.}
  \label{fig:TID}
  \end{subfigure}\hfill%
  \begin{subfigure}{.475\textwidth}
    \centering
    \includegraphics[width=1\textwidth,clip=]{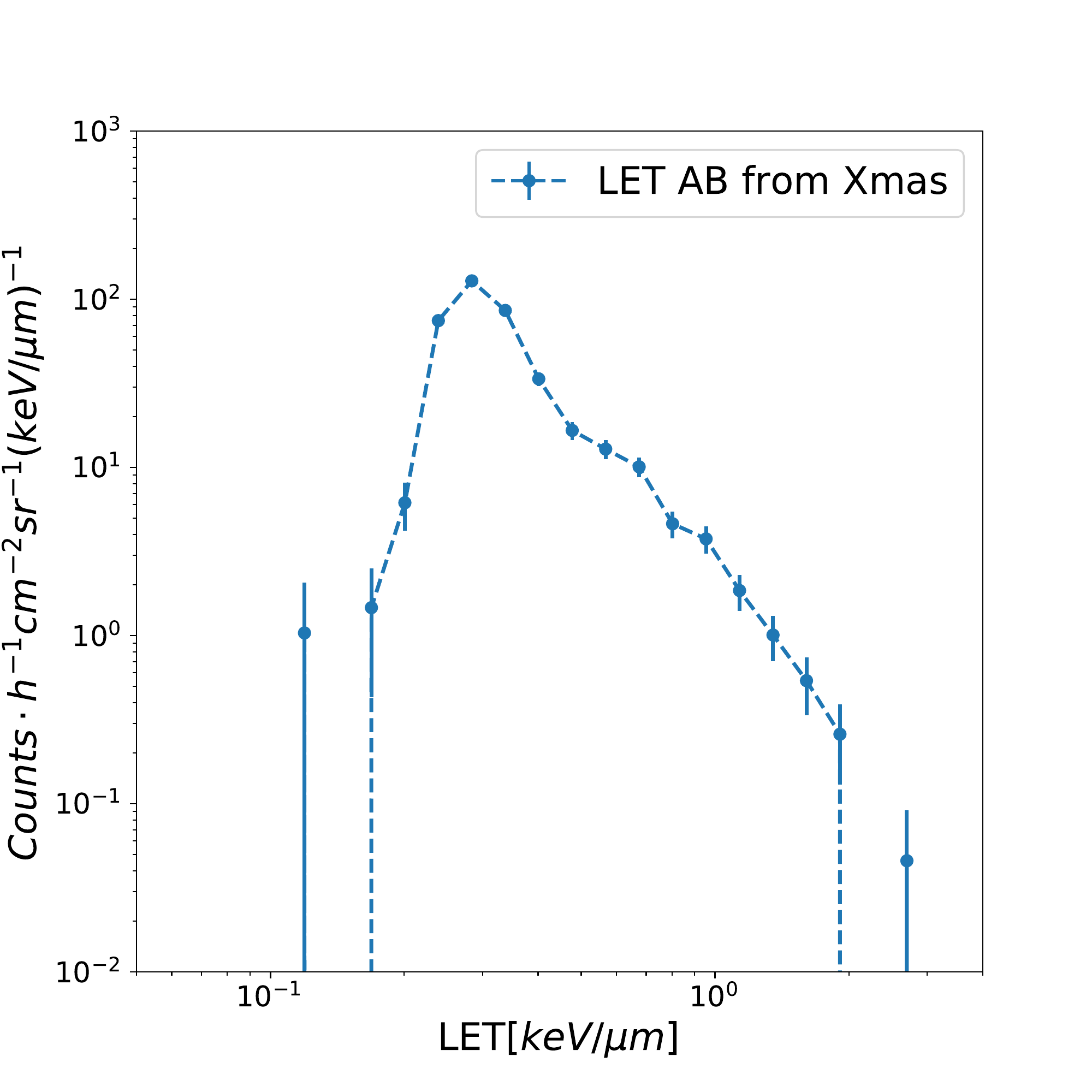}
    \caption{\acs{LET} spectrum acquired with \acs{LND} standing upright in our laboratory, pointing at the zenith. The energy deposit of minimally-ionizing muons is clearly seen around 0.3~keV/micron. Particles with lower primary energy contribute to higher LETs. The data shown correspond to the LET spectrum described in line 5 of Tab.~\ref{tab:LET}.
    }
    \label{fig:LET}
  \end{subfigure}
\caption{\acs{TID} (left) and \acs{LET} (right) acquired by \acs{LND} prior to re-delivery to NSSC in February 2018. Higher energy deposits than those visible are expected for heavy ions (\acs{GCR}s), but these are only very rarely seen on Earth at sea level (i.e., in our labs in Kiel). }
\label{fig:TID-LET}
\end{figure}

In order to derive the final, physical value of \acs{TID}, the value calculated using eq.~\ref{eq:TID2} needs to be converted into Joules and divided by the detector's mass, $m = 1.04$g,
\begin{equation}
  {\rm TID}_{\rm final} = 1.602\cdot 10^{-13} \cdot {\rm TID} / m,
  \label{eq:TID4}
\end{equation}
where \acs{TID} from eq.~\ref{eq:TID2} is expressed in keV and $m$ in grams, and \acs{TID}$_{\rm final}$ in Gy (J/kg). A coarse value for \acs{TID} is also given in the \acs{LND} 1-minute data, see Sec.~\ref{sec:1-min}. An example of a \acs{TID} spectrum is shown in Fig.~\ref{fig:TID} and was acquired by \acs{LND} in February 2018 during pre-delivery tests. The threshold of $\sim 40$ keV is clearly seen, as are the cosmic muons between 100 and 200 keV. The low energy deposits below that of cosmic muons are due to X-rays in the laboratory which are created by the interaction of high-energy \acs{GCR} products with the walls, floor, and ceiling of the laboratory, as well as the small background activity.

\subsubsection{Linear Energy Transfer spectra}

The Xmas plot contains three LET spectra defined by the detector combinations given in the three first lines of Tab.~\ref{tab:LET}. Their locations in the Xmas plot are given in Tab.~\ref{tab:xmas1}.
\begin{table}
  \centering
  \begin{tabular}{|clrrrrr|}\hline\hline
    & {\bf Det.} & {\bf \# bins} & {\bf $\braket{L}$ [$\mu$m]} & {\bf $\hat{L}$[$\mu$m]}& {\bf var($L$) [$\mu$m]} & {\bf g [mm$^2$sr]} \\\hline
1& $A\cdot B\cdot C\cdot \bar{I}$   \phantom{\Large Hi}       &  64 &     526   &   527 &   367    &     64     \\ 
2& $\bar{A}\cdot B\cdot I\cdot J$          &  64 &     619   &   593 &   15055  &     2019   \\
3& $A\cdot B\cdot I\cdot J$           &  $8 \times 64$ &     514   &   515 &   962    &     158    \\\hline
4& $2 \cup 3 \approx B\cdot J$    &  64 &     611   &   584 &   14774  &     2177   \\
5& $1 \cup 3 \approx A\cdot B$    &  64 &     517   &   517 &   823    &     222    \\\hline\hline
  \end{tabular}
  \caption{Properties of the various \acs{LET} spectra. The first three lines are the measurements performed by \acs{LND} while the following two lines are more relevant from a radiation protection point of view, as discussed in the text. Det.: detector coincidence used to define the LET. $\braket{L}$: mean path length through detector, $\hat{L}$: median of path length, var($L$): variance of pat length, $g$: $4 \pi$ geometric factor calculated with GEANT4. $2\cup 3$ and $1\cup 3$ in rows 4 and 5 indicate the logical union of conditions 2 and 3 ($(\bar{A}\cdot B\cdot I\cdot J) \cup (A\cdot B\cdot I\cdot J)$) and of 1 and 3 ($(A\cdot B\cdot C\cdot \bar{I}) \cup (A\cdot B\cdot I\cdot J)$).}
  \label{tab:LET}
\end{table}
The first three lines in Tab.~\ref{tab:LET} define the relevant quantities measured by \acs{LND}. The combinations in lines 4 and 5 define two LET spectra which are more relevant from a radiation protection point of view. Line 3 gives the narrowest viewcone and highest LET resolution, while line 4 gives a slightly wider viewcone with somewhat higher count rate. The combination given in line 5, finally, gives a very broad opening angle which is very similar to that of the Dostel instrument which has been operating on the International Space Station (ISS) for more than a decade \cite{berger-etal-2016,berger-etal-2017}. A low-resolution version (only 8 bins) of the LET spectrum 2 is sent to telemetry every minute, thus providing high time resolution LET measurements. An example for an LET spectrum is shown in Fig. ~\ref{fig:LET} and was acquired in February 2018 during pre-delivery test. It is based on high-resolution data from the Xmas plot. 


The energy for the \acs{LET} is measured in the B detector and is converted to bin number as given in eq.~\ref{eq:LET},
\begin{equation}
  {\rm bin \#} = 4\cdot \log_2(E_B/20),
  \label{eq:LET}
\end{equation}
where energies ($E_B$ and 20) are again measured in keV.
Because a high \acs{LET} can be due to either heavy ions or to stopping or nearly stopping ions, \acs{LND} gives one \acs{LET} spectrum which allows to discriminate between these two types of energy deposits in a 2-dimensional \acs{LET} histogram. It is defined by 
\begin{equation}
  {\rm bin \#}_x = 2\cdot \log_2(E_I/E_A) + 4, \quad \quad {\rm bin \#}_y = 4\cdot \log_2(E_B/20),
  \label{eq:LET2}
\end{equation}
where energies are again measured in keV. As can be seen in eq.~\ref{eq:LET2}, it uses the energy deposits in the A and I detectors in addition to the energy deposit in B, so it uses the detector combination with the smallest geometry factor and smallest path length dispersion, and thus highest resolution. The bins in the $x$ direction are defined completely analogously to the $x$ axis for the penetrating charged particles which are described in Sec.~\ref{sec:penetrating}.

To determine the \acs{LET} spectra, one needs to divide by the average path lengths, $\braket{L}$, given in Tab.~\ref{tab:LET}. The $x$-axis of the \acs{LET} spectrum is thus given by
\begin{equation}
  x_i = E_i/\braket{L} = 20 \cdot 2^{i/4}/\braket{L}\ \mbox{[keV/$\mu$m]}, \quad \quad y_i = \mbox{counts}_i
  \label{eq:LET3}
\end{equation}

\subsection{Charged Particles -- Ions}
\label{sec:charged}

\acs{LND} identifies charged particles by measuring the energy deposition in different combinations of its \acs{SSDs}. The mean energy loss per unit path length of a charged projectile particle with nuclear charge $Z_1$ and velocity $v$ in a detector is given by the well-known Bethe-Bloch equation \cite{bethe,bloch},
\begin{equation}
  \frac{\dd E}{\dd x} = - \frac{Z_1^2 e^4 n_e}{4 \pi \cdot \varepsilon_0^2 \beta^2 c^2 m_e} \cdot \left[ \ln\left(\frac{2 m_e  \beta^2 c^2}{\braket{E_B}}\right) - \ln(1 - \beta^2) - \beta^2  \right], 
  \label{eq:BB}
\end{equation}
where $\beta = v/c$, $c$ the speed of light, $E_B$ is the ionization energy of the target (173 eV for Si), and $n_e$ is the electron density in the target ($6.7 \times 10^{29}$m$^{-3}$ for Si), $m_e$ the electron mass, and $\varepsilon_0$ the permittivity of free space. The expression in the square brackets simplifies to a constant plus a slowly changing  $\ln(\beta^2)$ in the energy range covered by \acs{LND}. Thus, we may approximate the energy loss in the front most detector, A, as
\begin{equation}
  E_A \propto \frac{Z_1^2 m_1}{E_{\rm tot}},
  \label{eq:BB2}
\end{equation}
where $E_{\rm tot}$ is the measured initial energy of the stopping projectile particle, and $E_A$ is the energy deposited in the front detector (A).
The product of the total energy and the energy deposited in A (eq.~\ref{eq:BB2}) is then given by
\begin{equation}
  Y \doteq E_{\rm tot} \cdot E_A \propto E_{\rm tot}\cdot \frac{Z_1^2 m_1}{E_{\rm tot}} \propto Z_1^2 m_1 \ \approx \ {\rm const.}
  \label{eq:xmasy}
\end{equation}
This quantity, $Y$, only depends on the particle species, i.e, its nuclear charge, $Z_1$, and mass, $m$. Except for some very weak dependencies on $\ln(\beta^2)$ all other quantities have canceled out. Thus this quantity only depends on the particle's properties which allows us to discriminate even some isotopes of the various elements. $Y$ as defined in eq.~\ref{eq:xmasy} is the quantity plotted as the $y$-axis of the Xmas plot for particles which stop in detectors B, C, and D. Stopping particles up to a penetration depth of detector D are identified by the condition that there is no energy deposited in detector E. Because the following detectors, F, G, H, I, and J all have non-sensitive material in front of them, the analogous quantity can not be used in the same manner. For them, the quantities given in Tab.~\ref{tab:axis-defs} are plotted along the $y$-axis of the Xmas plot.

The quantity plotted along the $x$-axis of the Xmas plot is given by the ratio of the the total deposited energy, $E_{\rm tot}$, and the energy measured in the front detector, $E_A$,
\begin{equation}
X \doteq E_{\rm tot} / E_A.
    \label{eq:xmasx}
\end{equation}
Using the approximation of eq.~\ref{eq:BB2} this ratio is proportional to $E_{\rm tot}^2/(Z_1^2\cdot m)$, i.e., is sensitive to the total (primary) energy of the stopping particle. 

For stopping particles, the quantities for the  $x$ and $y$-axis are summarized in Tab.~\ref{tab:axis-defs}. As indicated by red counters in Fig.~\ref{fig:xmas2} the combination $(X,Y)$ identifies individual elements or element groups for all particles stopping in detectors B through I. The grey-shaded strip marked by "Stop in B" has two such counters for protons and three each for both He isotopes. Particles with higher total energy tend to lie at larger values of $X \propto E^2_{\rm tot}$. The neighboring strip marked by "Stop in C" shows the same kind of data for particles stopping in C, etc.

\begin{table}
  \centering
  \begin{tabular}{|lc|}\hline\hline
    {\bf Det.} & {\bf $x$-axis} \\\hline
    stopping in B & $8\cdot\log_2((E_A + E_B) / E_A)$ \\
    stopping in C & $8\cdot\log_2((E_A + E_B + E_C) / E_A) - 8$  \\
    stopping in D & $16\cdot\log_2((E_A + E_B + E_C + E_{D1}) / E_A) - 24$ \\
    stopping in E & $16\cdot\log_2((E_A + E_B + E_C + E_{D1}) / E_A) - 24$  \\
    stopping in F & $16\cdot\log_2((E_A + E_B + E_C + E_{D1}) / E_A) - 24$  \\
    stopping in G & $16\cdot\log_2((E_A + E_B + E_C + E_{D1}) / E_A) - 24$  \\
    stopping in H & $16\cdot\log_2((E_A + E_B + E_C + E_{D1}) / E_A) - 24$ \\ 
    stopping in I & $2\cdot\log_2(E_I / E_A) + 8$ \\\hline 
    {\bf Det.}    & {\bf $y$-axis} \\\hline
    stopping in B & $4\cdot\log_2((E_A + E_B) \cdot E_A / (8000 \cdot 4000))$ \\
    stopping in C & $4\cdot\log_2((E_A + E_B +E_C) \cdot E_A / (10000 \cdot 4000)) + 4$\\
    stopping in D & $4\cdot\log_2((E_A + E_B +E_C + E_{D1}) \cdot E_A / (1200 \cdot 6000)) - 8$ \\
    stopping in E & $4\cdot\log_2((E_A + E_B +E_C + E_{D1}) \cdot E_A) / (1200 \cdot 6000)) + 2$\\
    stopping in F & $4\cdot\log_2((E_A + E_B +E_C + E_{D1}) \cdot E_A) / (1200 \cdot 6000)) + 7$\\
    stopping in G & $4\cdot\log_2((E_A + E_B +E_C + E_{D1}) \cdot E_A) / (1200 \cdot 6000)) + 7$ \\
    stopping in H & $4\cdot\log_2((E_A + E_B +E_C + E_{D1}) \cdot E_A) / (1200 \cdot 6000)) + 8$ \\
    stopping in I & $8\cdot\log_2((E_A + E_B +E_C + E_D + E_I)/4000)$ \\\hline
    \multicolumn{2}{|l|}{\bf Penetrating Particles}  \\\hline
     {\bf $x$-axis}   & $4 \cdot \log_2 (E_i/E_A) + 16$ \\
     {\bf $y$-axis}   & $4 \cdot \log_2 ((E_{B1} + E_C + E_D)/100)$ \\\hline\hline
  \end{tabular}
  \caption{Quantities used as the $x$ and $y$-axis for the stopping charged-particle regions of the Xmas plot. Det: Detector in which the particle stops. The last line gives the quantities used for penetrating particles (column J in the Xmas plot).}
  \label{tab:axis-defs}
\end{table}

Energetic electrons need to be treated separately because of their high penetration power. A similar identification scheme is used for them, but their resulting memory addresses ($(X,Y)$ coordinates) in the Xmas plot are given in Tab.~\ref{tab:e-electrons}, as explained in section~\ref{sec:electrons}.

\subsubsection{Energy Ranges for Stopping Ions}

As particles enter the front foil of \acs{LND} and penetrate the A detector they loose energy. For a particle to be registered, it must trigger the B detector which gives the minimum energy it must have in order to be measured by \acs{LND}. If it has more energy, it may penetrate the B detector and stop in C, etc. The stopping energies are given in Tab.~\ref{tab:stop-energies}. They are based on \acs{GEANT4} simulations of the \acs{LND} sensor and include all relevant layers such as the Gd-absorber, its surrounding Al-housing, etc. The simulations were performed for an isotropic source on top of the \acs{LND} entrance foil. 
\begin{table}
  \centering
  \begin{tabular}{|lrrrrr|}\hline\hline
Stop in       & proton      & $^3$He             & $^4$He                     & CNO           & Heavy Ions\\
              & [MeV]       & [MeV/nuc]          & [MeV/nuc]                  & [MeV/nuc]     & [MeV/nuc] \\\hline
B dps$_1$     & {9.0-10.6}  & {10.6-11.4}        & {8.9-9.5}                  &   {16.6-20.5} &  {24.6-37.7} \\
B dps$_2$     & {10.7-12.7} & {11.4-12.6}        & {9.6-10.9}                 &   {20.5-23.5} &  {31.0-46.4} \\
B dps$_3$     &  -          & {12.9-14.8}        & {10.9-12.6}                &   {-} 		  &  {-}         \\
C             & {12.8-15.7} & {15.0-18.5}        & {12.8-15.7}                &   {25.5-29.3} &  {38.2-58.5} \\
D             & {15.9-18.4} &{18.7-21.7}&{15.9-18.5}&   {31.0-36.4} &  {47.5-68.7} \\
E             & {18.6-21.0} &{21.7-24.6}$^a$ &{18.5-21.0}$^a$ &   {37.3-38.6} &  {52.1-78.9} \\
F             & {21.2-29.2} &{24.9-34.4}&{21.0-29.3}&   {42.3-59.1} &  {67.1-105.2} \\
G             & {29.6-31.3} &{34.8-36.4}&{29.3-31.4}&   {54.5-62.6} &  {91.6-122.2} \\
H             & {31.5-33.0} &{36.9-38.6}&{31.4-32.8}&   {57.8-66.4} &  {79.8-131.0} \\
I             & {33.4-34.5} &{39.0-40.4}&{32.8-34.4}&   {61.2-72.8} &  {101.6-132.5} \\\hline\hline
{Stop in}     & {proton}    &{$^3$He}            & {$^4$He}           &  {CNO}      &   {Heavy Ions} \\\hline
{B dps$_1$}   &     {H1}    &{3He1}              & {4He1}             &  {CNO1}     &  {ions1}\\
{B dps$_2$}   &     {H2}    &{3He2}              & {4He2}             &  {CNO2}     &  {ions2}\\
{B dps$_3$}   &     {-}     &{3He3}              & {4He3}             &  {-}        &  {-}    \\
{C}           &     {H3}    &{3He4}              & {4He4}             &  {CNO3}     &  {ions3}\\
{D}           &     {H4}    &{3He5}              & {4He5}             &  {CNO4}     &  {ions4}\\
{E}           &     {H5}    &{4He6}$^a$              & {4He6}$^a$             &  {CNO5}     &  {ions5}\\
{F}           &     {H6}    &{4He7}              & {4He7}             &  {CNO6}     &  {ions6}\\
{G}           &     {H7}    &{4He8}              & {4He8}             &  {CNO7}     &  {ions7}\\
{H}           &     {H8}    &{4He9}              & {4He9}             &  {CNO8}     &  {ions8}\\
{I}           &     {H9}    &{4He10}             &{4He10}             &  {CNO9}     & {ions9}\\\hline\hline
Penetrating   &        &                &                &         &    \\\hline
pene1         &             &{56.5-182.9}        & {48.6-180.8}       &             &    \\
pene2         &             &                    &                    &{$\geq$88.5}$^{b}$& {$\geq$150.4}$^{b}$ \\
pene3         &             &                    &                    & {95.9-147.0}&{$\geq$111.4}$^{b}$\\
pene4         &             &{40.7-56.5}         &{34.8-45.9}         &             &                \\
pene5         &             &                    &                    & {84.5-91.6} &                \\
pene6         &             &                    &                    & {70.3-93.8} &  {78.9-106.4}    \\
pene7         &             &                    &                    & {75.3-78.9} &  {105.2-170.7}   \\
H10           &{$\geq$156.2}&                    &                    &             &                \\
H11           & {42.3-139.2}&{$\geq$198.3}       &{$\geq$198.3}       &             &                \\
H12           &             &                    &                    &             &                \\
H13           &             &                    &                    &             &                \\
H14           &{34.9-40.8}  &                    &                    &             & \\\hline\hline

  \end{tabular}
  \caption{List of primary energy ranges that protons, $^3$He, $^4$He, CNO, and heavy ions need to have to stop in the \acs{LND} detectors. ``B dps$_i$'', ``pene $i$'' and H10 - H14 refer to the multiple red counters for protons (ions) in the Xmas plot, Fig.~\ref{fig:xmas2}, which are discussed in sections~\ref{sec:10-min} and \ref{sec:1-min}.
    Example: A proton must have a primary energy $16.0 < E < 18.7$ MeV so that it will stop in D. The lower half of the table shows the name of the corresponding energy bins in the upper half of this table. The positions of these energy bins in the X-mas plot are indicated with white labels in Figure \ref{tab:stop-energies}. The lowest part of the table gives the minimum energy needed to stop in or penetrate J. \\$^a$ $^3$He and $^4$He share the dps boxes from detectors E to I
    \\$^b$ 10\% of maximum used instead of 50\% maximum
    }
  \label{tab:stop-energies}
\end{table} 

\subsection{Penetrating Charged Particles}
\label{sec:penetrating}

The spectrum of energetic particles, especially the \acs{GCR}, has a substantial flux of particles with energies which exceed the stopping power of \acs{LND}. They are measured in the penetrating channel of \acs{LND} which is defined by particles which penetrate all detectors, and stop in or penetrate the J detector. LND can not discriminate particles stopping in the last detector, J, from particles that penetrate it because J serves as the anti-coincidence detector for particles stopping in LND.  Most of these penetrating particles are close to minimally ionizing, i.e., they have a nearly constant energy loss across \acs{LND}. Barely penetrating particles have primary energies which result in smaller energy losses in the front detectors A, B, C, D, than in the rear detectors I and J. To discriminate barely stopping particles from  penetrating particles, the Xmas plot is redefined as a $32 \times 64$ matrix with its $X$ and $Y$ axis coordinates computed according to the rows marked "Penetrating Particles" in Tab.~\ref{tab:axis-defs}. This is the definition of the rightmost panel of the Xmas plot marked as "Penetrate J". One easily recognizes the fully relativistic particles as a nearly straight line from the bottom to the top and the nearly stopping particles which turn to the right, i.e., towards larger values of $4\log_2(E_I/E_A)$. Particles which veer to the left move upwards through \acs{LND}, i.e., away from the Moon. They are part of the secondary particle population which is created by the interaction of (mainly) GCRs with the lunar soil. Penetrating electrons lie at the bottom of this panel and are clearly separated from barely penetrating protons, but not from fully relativistic protons which is as expected.


\subsection{Charged Particles -- Electrons}
\label{sec:electrons}

The calculation of the $(X,Y)$ addresses for electrons in the Xmas plot results in vanishing $Y$ values and so most electron data products are plotted along the $X$ axis ($Y=0$) of Fig.~\ref{fig:xmas2}. Because electrons are abundant, there are also one- and ten-minute electron data products. These are indicated in Fig.~\ref{fig:xmas2} by E, F, GHI, and E1,\dots E4, \ldots H1, \ldots H4. The primary energy ranges corresponding to these one- and ten-minute data products are given in Tab.~\ref{tab:e-electrons}. Electrons stopping in the B, C, and D detectors are mapped to the two right-most columns of the Xmas plot, columns 273 and 274. Those addresses are calculated as {16}$\cdot \log_2(T)$, where $T$ is the total deposited energy (measured in keV). These predominantly low-energy electrons have four one-minute data products indicated by dps1 -- dps4 in Fig.~\ref{fig:xmas2}, their ranges of {\em deposited} energy is also given in Tab.~\ref{tab:e-electrons}.



\begin{table}
  \centering
  \begin{tabular}{|lrrrrr|}\hline\hline
     {\bf Electron channel} & \multicolumn{2}{c}{\bf bin edges}   & {\bf $E_{\rm dep}$} & & {\bf time res.}\\
                            & {\bf low}          & {\bf high}        & {\bf [MeV]} & & {\bf [minutes]}\\\hline
  {\bf Column {273}}    & \multicolumn{2}{c}{\bf Row \#s}     &             & & \\\hline
                 $e^-$ dps$_1$   & 10                &  {17}            & {0.5-0.707} & & 1\\
                 $e^-$ dps$_2$   & 18                &  {25}            & {0.707-1} & & 1\\
                 $e^-$ dps$_3$   & 26                &  {33}        & {1-1.414} & & 1\\
                 $e^-$ dps$_4$   & 34                &  {41}        & {1.414-2} & & 1\\\hline
  {\bf            }    & \multicolumn{2}{c}{\bf }  &\multicolumn{2}{c}{\bf $E_{\rm prim}$ [MeV]}       &   \\
  {\bf Row 0           }    & \multicolumn{2}{c}{\bf Column \#s}  &{\bf half max} & {\bf 20\% -- 20\%}    &        \\\hline \hline

                    {Xmas-E} & {97}           & {126}         &  {1.3-2.7}   & {1.2-4.3}  & 1\\
				    {Xmas-F} & {129}          & {158}         &  {1.9-5.2}   & {1.6-8.5}  & 1\\
				    {Xmas-G$\cup$H$\cup$I}  &{160} & {239}    &  {3.0-9.0}   & {2.7-16.3} & 1\\  

\hline	\hline				   
                       Xmas-E1  & {96}              &  {96}    & {1.5-4.2} & {1.3-9.8} & 10\\
                       Xmas-E2  & {97}              &  {102}   & {1.3-2.8} & {1.2-5.4} & 10\\
                       Xmas-E3  & {118}             &  {126}   & {1.5-2.8} & {1.3-4.3} &10\\
                       Xmas-E4  & {127}             & {127}    & {2.0-3.4} & {1.8-4.9} &10\\                       
                       Xmas-F1  & {128}             &  {128}   & {2.2-6.7} & {1.8-12.3} & 10\\
                       Xmas-F2  & {129}             &  {132}   & {2.0-5.3} & {1.7-9.3} & 10\\
                       Xmas-F3  & {145}             &  {158}   & {2.1-5.7} & {1.8-8.9} & 10\\
                       Xmas-F4  & {159}             &  {159}   & {3.0-6.2} & {2.4-8.5} & 10\\
                       Xmas-G1  & {160}             &  {160}   & {3.1-7.2} & {2.7-12.9} & 10\\
                       Xmas-G2  & {161}             &  {165}   & {2.8-6.5} & {2.6-10.7} & 10\\
                       Xmas-G3  & {166}             &  {172}   & {2.8-6.1} & {2.6-10.2} & 10\\
                       Xmas-G4  & {173}             &  {191}   & {2.9-7.5} & {2.7-9.8} & 10\\
                       Xmas-H1  & {192}             &  {192}   & {3.6-16.1} & {2.9-29.7} & 10\\
                       Xmas-H2  & {193}             &  {197}   & {3.3-12.5} & {2.8-24.7} & 10\\
                       Xmas-H3  & {198}             &  {204}   & {3.1-12.2} & {2.8-20.5} & 10\\
                       Xmas-H4  & {205}             &  {223}   & {3.3-11.6} & {2.9-22.5} & 10\\
\hline\hline
  \end{tabular}
  \caption{Deposited and primary energy ranges ($E_{\rm dep}$ and $E_{\rm prim}$) for electrons measured in LND detectors and/or dps counters/counters. Note that these show very large variability as electrons scatter a lot. The column half max gives the \acs{FWHM} range around the modeled maximum, the following column gives the range between $\pm$ 20\% of the maximum. The positions of these energy bins in the Xmas plot are indicated with black labels in Figure~\ref{fig:xmas2}. For instance, Xmas-E3 refers to the electron region marked by E3 in the Xmas plot (rows 118 - 126, column 0). The last column gives the time resolution at which these data products are put into telemetry.}
  \label{tab:e-electrons}
\end{table}

\begin{figure}
   \centering
  \begin{subfigure}{.475\textwidth}
  \centering
  \includegraphics[width=0.9\textwidth,clip=]{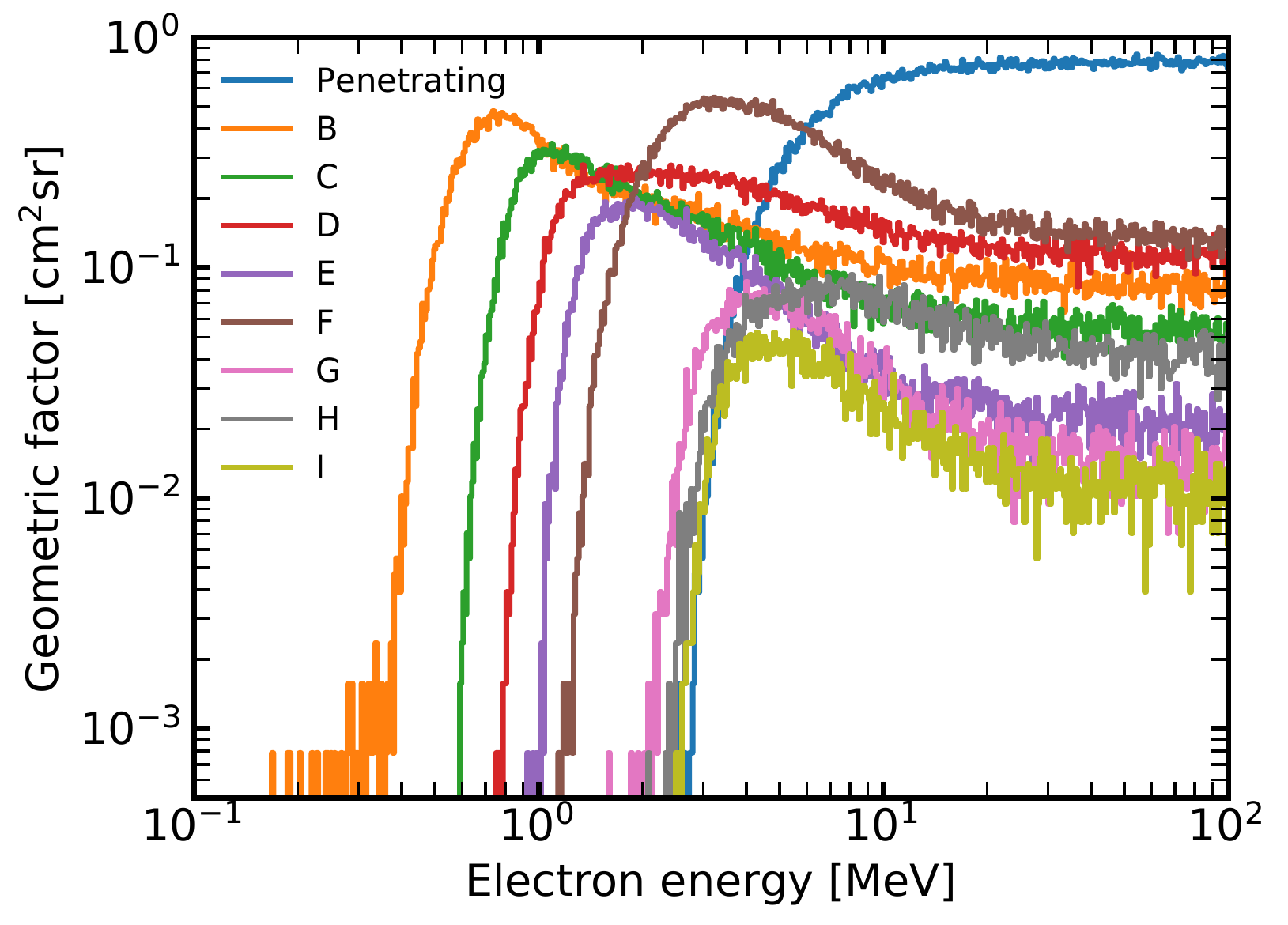}
  \caption{These simulated count spectra show that electrons need at least 300 keV to be identified in the B, while the probability to be seen in this detector peaks at approximately 0.7 MeV. This peak depends on the incident spectrum, of course. ``Deeper'' detectors have higher detection thresholds, allowing for a coarse identification of the spectral shape of electrons. }
  \label{fig:e-B-I}
  \end{subfigure}\hfill%
  \begin{subfigure}{.475\textwidth}
    \centering
    \includegraphics[width=0.9\textwidth,clip=]{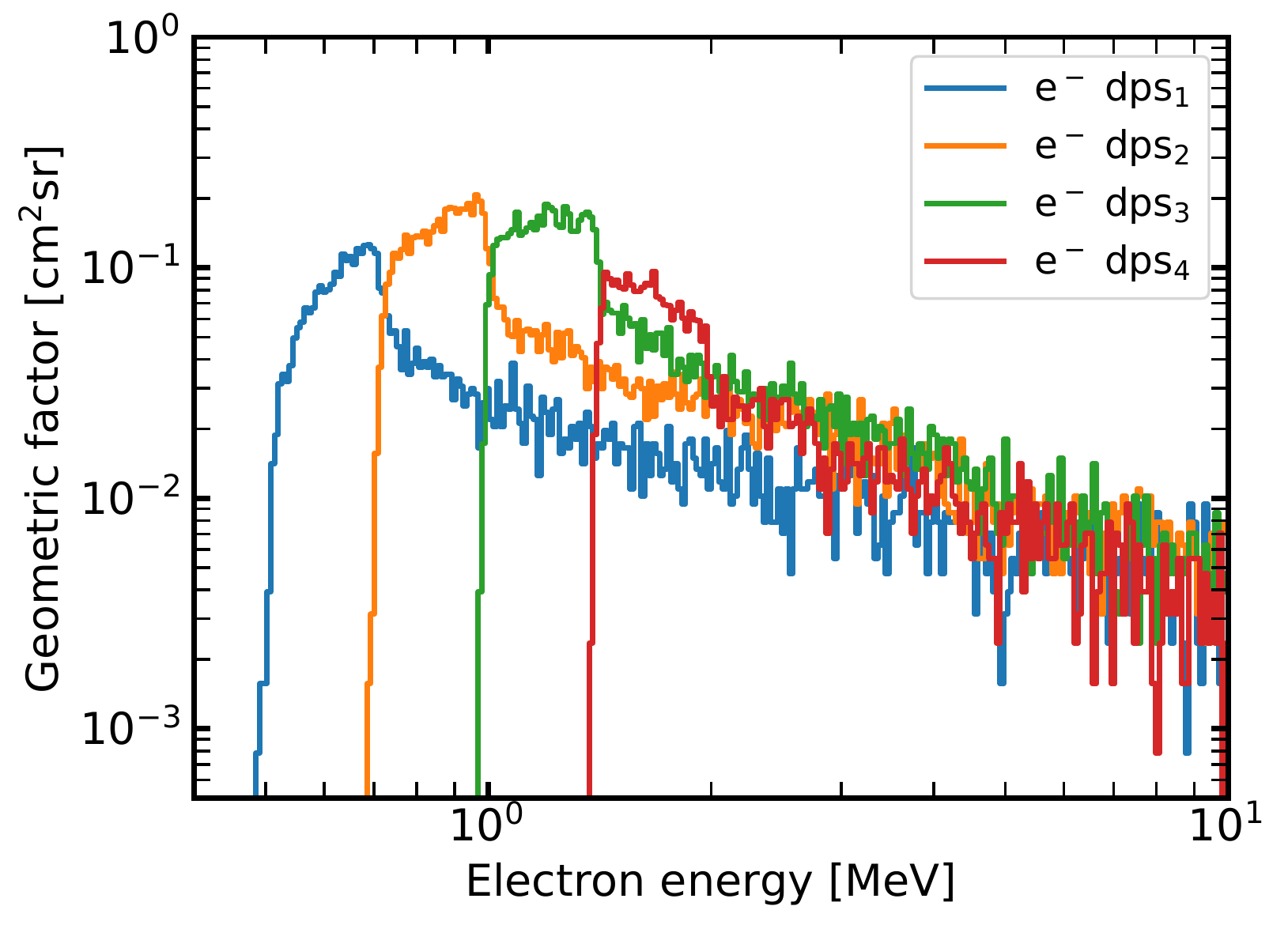}
    \caption{These simulated count spectra show that electrons need at least 0.5 MeV to be identified in the $e^{-}$ dps$_1$ data product, while the probability to be seen in this data product peaks at approximately 0.7 MeV. This dps ``threshold'' increases with increasing primary energy which allows us to give a very coarse estimate of the electron spectrum, see also Fig.~\ref{fig:e-B-I}.}
    \label{fig:e-dps1234}
  \end{subfigure}
\caption{Electron penetration depths inside \acs{LND} give information about their primary energy. The mapping is not 1:1 though, and an inversion process as described by eq.~\ref{eq:fredholm} will be needed to obtain higher fidelity electron spectra. 
}
\label{fig:electrons}
\end{figure}

The increasing penetration depth of electrons with increasing primary energy can be used for an initial, coarse estimate of the electron spectrum. Figure~\ref{fig:electrons} shows how this works. In the left-hand panel, Fig.~\ref{fig:e-B-I}, the simulated counts are shown for an isotropic, $E^{-1}$ incident electron spectrum. One easily sees that increasing primary energy of electrons results in increasing penetration depth into the detector stack. electrons need to have a kinetic energy of at last 300 keV to trigger the B detector. The results for the F detector (and to a lesser degree the H detector) appears to ``stick out'' of the which is also expected, because the trigger condition is $B\cdot C\cdot D \cdot E\cdot F\cdot \bar{G}$ and there is the Al-Gd-Al sandwich between the F and G detectors. The blue curve shows results for penetrating electrons with primary energies exceeding 3 MeV. The right-hand panel (Fig.~\ref{fig:e-dps1234}) shows the response of the dps1 -- dps4 bins of the Xmas plot to the same electron spectrum. It is dominated by electrons in the primary energy range $0.5 < E < 2$ MeV. 
These four data products are part of the \acs{LND} 1-minute data, as are those marked E, F, and GHI in the Xmas plot. 


Note that determining the ``true'' electron spectrum will require an inversion approach which accounts for the \acs{LND} instrument response function, $K_{e^-}(E_{\rm det},E_{\rm prim})$, which gives the geometry factors for an electron of primary energy $E_{\rm prim}$ to be measured at an energy $E_{\rm det}$. This can be achieved using constrained optimization or other optimization techniques \cite{boehm-etal-2007,kharytonovetal2009} which solve the Fredholm integral given in eq.~\ref{eq:fredholm}.
\begin{equation}
  \vec{z}(E_{\rm det}) = \int_0^\infty K_{e^-}(E_{\rm det},E_{\rm prim}) f(E_{\rm prim}) \dd E_{\rm prim},
  \label{eq:fredholm}
\end{equation}
where $f(E_{\rm prim})$ is the primary spectrum and $\vec{z}(E_{\rm det})$ are the measured counts in the energy bins $E_{\rm prim}$. 

\subsection{1-Hour Data Products}
\label{sec:1-hour}





The Xmas plot (Fig.~\ref{fig:xmas2}) is LND's principal data product and is accumulated over the course of an hour. As explained in Sec.~\ref{sec:xmas}, there are two Xmas plots, one active (i.e., being accumulated) and one inactive (i.e., being sent to telemetry). 
It contains the high-resolution data needed to discriminate between different particle species, especially the rare heavy ions. The data products contained in the Xmas plot have been discussed in sections \ref{sec:xmas} -- \ref{sec:electrons}.

\subsection{10-Minute Data Products}
\label{sec:10-min}

To allow for higher time resolution, LND reads out certain regions in the active Xmas plot more often than once an hour. These regions are shown as red boxes in Fig.~\ref{fig:xmas2} and we refer to them as boxes or counters in the following. Because the number of counts in an individual ``pixel'' in the Xmas plot will necessarily be small, these large 10-minute boxes contain many Xmas plot counters to allow us to obtain statistically significant numbers, at least during large solar particle events. In the case of particles stopping in detector E, we see that a box is centered on pixel 112 in the $x$ direction and pixel 20 in the $y$ direction and is marked "4He6". This region in the Xmas plot corresponds to $^4$He particles stopping in E. Thus, we know that this particle stopped in E and that it was a $^4$He particle. Since it takes a specific energy  to penetrate the detectors prior to E and to stop in E (see Tab.~\ref{tab:stop-energies}) we also know the primary energy of the $^4$He particles that stopped in E. This then gives us a point in the differential flux of $^4$He particles. 

The red box which we have placed around the $^4$He particles stopping in E also serves as a counter in LND. The LND processing scheduler ensures that the sum of all the counters in the individual pixels in this box is determined every 10 minutes. This sum is then put into the telemetry. Note that this counter now contains the number of counts accumulated in this box since the Xmas plot was last initialized. The number of counts in this box are thus cumulative for every 10 minute interval in an hour. The total number of counts in the past 10 minutes is thus determined by computing (on ground) the difference between the latest counter and the preceding one. These differences have already been computed in the published data. These 10-minute ``counters'' are 24-bit integer counters, they are not protected against \acs{SEU}s since these are unlikely to happen in any given 10-minute interval. No compression is performed on the counters. The data products available at 10-minute time resolution are listed in Tab.~\ref{tab:data}. 


The energy ranges for the various counter boxes are given in Tab.~\ref{tab:stop-energies}. Their locations can be seen in Fig.~\ref{fig:xmas2} and their meanings are obvious from it and Tab.~\ref{tab:stop-energies}.

\subsection{1-Minute Data Products}
\label{sec:1-min}

\acs{LND} also provides data at a time resolution of 1 minute. The logics are exactly the same as for the 10-minute data, counters are accumulated in boxes and read out once every minute. They are cumulative, i.e., one needs to take the difference between the current value and the preceding one to obtain the count rate. The data products available at 1-minute time resolution are listed in Tab.~\ref{tab:data}. The energy ranges for the various ions are also given in Tab.~\ref{tab:stop-energies}, the positions of these one-minute counter boxes are shown in Fig.~\ref{fig:xmas2}.

\subsection{Dead-time corrections}
\label{sec:deadtimes}

The accumulation of a Xmas plot happens during a full hour, but is stopped for 500 $\mu$s every second. Thus, the accumulation time of the Xmas plot is not 3600 seconds, but 3598.2 seconds. This is a small correction, but mentioned here for completeness sake. Similarly, the \underline{first} dps$_1$ data product in every hour is only accumulated for 59 seconds, all others for 60 seconds per minute, and, the \underline{first} dps$_{10}$ data product in one hour is only accumulated for 599 seconds, all others for 600 seconds every 10 minutes. Corrections for these small discrepancies are applied to the published data.

In addition to the correct accumulation times of the different data products, some dead time corrections need to be applied to LND's data. An approximation of the dead time that occurs while LND reads out its detectors can be calculated by multiplying the number of detector readouts by 7 $\mu$s. The number of readouts can be taken from the L2 trigger counts, but it needs to be considered which L2 triggers contribute to which data product. This is easiest to understand by inspecting Fig.~\ref{fig:L1L2L3_diagram}. The simplest situation is when LND detects a neutron in detector C1, as this affects only one single detector and trigger, L2[2]. There is also only one corresponding L3 trigger, L3[1]. The dead time is then calculated by multiplying the number of L2[2] triggers by a dead time of 7 $\mu$s. A more complicated dead time correction is that for TID because it gets contributions from L2[0] (LET) via the L3[3] (LET) trigger, but also the L2[1] (TID) and L2[7] (heavy ions) via the L3[0] trigger path to LND's TID data product (shown in blue in Fig.~\ref{fig:L1L2L3_diagram}). Thus, the dead time associated with TID is the sum of the counts in L2[0], L2[1], and L2[7] multiplied by 7 $\mu$s. Finally, the dead time correction for the thermal neutron channels in E, F, G and H is done using the sum of the trigger counts in L2[3] through L2[6]. The eight L2 counters are in telemetry at 1 minute resolution, so this correction can be made at LND's highest measuring cadence. For the data products sent at lower cadence, the L2 counters are summed up over the respective time intervals to calculate the corresponding dead time.

LND's accumulation times and dead times are summarized in Tab.~\ref{tab:deadtimes}. They need to be considered by correcting the nominal accumulation time, $\dd t$ (3600, 600 or 60 seconds, respectively), to
\begin{equation}
  \dd t_{\rm true} = \mbox{accumulation time} - \mbox{dead time},
  \label{eq:deadtime}
\end{equation}
where the accumulation time and dead time are taken from Tab.~\ref{tab:deadtimes}. Count rates are thus corrected for dead-time effects using the following expression,
\begin{equation}
{\rm true\ count\ rate} = {\rm measured\ count\ rate} \cdot \frac{\dd t}{\dd t_{\rm true}},
    \label{eq:dead-time-corr}
\end{equation}
where $\dd t$ is the nominal, uncorrected accumulation time for a data product. Dead time corrections have been applied to the published data.


\begin{table}
  \centering
  \begin{tabular}{|lr|}\hline\hline
    {\bf Type of data product} & {\bf Accumulation time [s]} \\\hline
    Xmas plot                                & 3598.2\\
    \underline{first} dps$_1$ per hour          & 59 \\
    other dps$_1$                               & 60 \\
    \underline{first} dps$_{10}$ per hour         & 599 \\
    other dps$_{10}$                              & 600 \\
    \hline
    {\bf Type of measurement} & {\bf Dead time [$\mu$s]} \\
    \hline
    TID, LET, charged particles & (L2[0] $+$ L2[1] $+$ L2[7])$\cdot 7$\\
    Neutrons in C1 & L2[2]$\cdot 7$\\
    Thermal neutrons in E, F, G, H & (L2[3] $+$ L2[4] $+$ L2[5] $+$ L2[6])$\cdot 7$\\
    \hline\hline
  \end{tabular}
  \caption{Accumulation times and dead times to be considered when producing \acs{LND} data products. The ``first'' are underlined because this applies only to the first dps packet per hour. Dead time corrections are discussed in Sec.~\ref{sec:deadtimes}. All counters associated with a data product given in Fig.~\ref{fig:L1L2L3_diagram} contribute to its dead time and thus their sum must be multiplied by 7 $\mu$s to obtain the dead time.}
  \label{tab:deadtimes}
\end{table}

\subsection{Calibration of \acs{LND}}
\label{sec:calib}

The \acs{LND} data products are provided in calibrated values, i.e., after all conversions from instrumental measurement units to the reported physical units have been applied. The calibration of \acs{LND} was was performed using the flight model (FM), the flight spare (FS) unit, and also the engineering model (EM). Because \acs{LND} provides very diverse measurements and data products, its calibration was an extended effort which included a number of facilities in Europe and China. These are summarized in Tab.~\ref{tab:calibration-overview} and will be published elsewhere in the literature. \acs{LND} itself was not calibrated with fast neutrons, but uses the calibration of the Flight Radiation Environment Detector (FRED, \cite{moeller-2013}) which also served as a prototype for the fast neutron detection capabilities of \acs{LND}.
\acs{FRED} was calibrated with neutrons at the Physikalisch-Technische Bundesanstalt (PTB) in Braunschweig (5 MeV $<E_n<$ 19 MeV) and with higher energy neutrons at iThemba in South Africa \cite{moeller-2013}. \acs{FRED} contains near-identical detectors B, C, D in the same stack configuration as \acs{LND}. The main difference lies in their thicknesses (300 $\mu$m in FRED and 500 $\mu$m in LND). These are accounted for by modeling the slightly different responses with GEANT4.

\begin{table}
  \centering
  \begin{tabular}{|llll|}\hline\hline
  \multicolumn{2}{|c}{\bf Calibration}& & \\ \cline{1-2}
    {\bf facility} & {\bf purpose} & {\bf LND Model} &{\bf Time frame} \\\hline
    CAU Kiel        & Bi-207 gammas             & FM, FS, EM& Summer 2017              \\
    CERN/CERF       & dose                      & FM        & 2017-06-17 -- 2017-06-18 \\
    HIMAC           & ions: H, He, C, O, Ar         & FM        & 2017-06-30 -- 2017-07-04 \\
    ATI             & thermal neutrons          & FM, FS    & 2017-09-25 -- 2017-09-28 \\
    18th Institute  & RTG \& RHU background     & FS        & 2018-07-09 -- 2018-07-12 \\
    NSSC            & electrons                 & EM        & 2018-08-17 -- 2018-08-20 \\ \hline\hline
  \end{tabular}
  \caption{Overview of calibrations performed with LND.}
  \label{tab:calibration-overview}
\end{table}




\begin{acknowledgements}
The Lunar Lander Neutron and Dosimetry (\acs{LND}) instrument is supported by the German Space Agency, DLR, and its Space Administration under grant 50 JR 1604 to the Christian-Albrechts-University (CAU) Kiel and supported by Beijing Municipal Science and Technology Commission, Grant No.Z181100002918003 to the National Space Science Center (NSSC). We would like to thank the following facilities for allowing us to calibrate LND: CERN/CERF in Geneva, Switzerland; HIMAC in Chiba, Japan; ATI in Vienna, Austria; SEF in NSSC, Beijing, China. We thank Oliver Angerer at DLR for his unwavering support of LND. We also thank the LESEC/CAST their joint effort for LND.
\end{acknowledgements}

\bibliographystyle{spphys}       
\bibliography{diss.bib}

%
%

\end{document}